\documentclass[10pt,letterpaper]{article}
\usepackage[top=0.85in,left=2.75in,footskip=0.75in]{geometry}

\usepackage{amsmath,amssymb}
\usepackage{braket}

\usepackage{changepage}

\usepackage{textcomp,marvosym}

\usepackage{cite}

\usepackage{nameref,hyperref}
\usepackage{subcaption}
\usepackage[graphicx]{realboxes}

\usepackage[right]{lineno}

\usepackage[nopatch=eqnum]{microtype}
\DisableLigatures[f]{encoding = *, family = * }

\usepackage[table]{xcolor}
\usepackage{booktabs}

\usepackage{array}

\newcolumntype{+}{!{\vrule width 2pt}}

\newlength\savedwidth

\raggedright
\usepackage[aboveskip=1pt,labelfont=bf,labelsep=period,justification=raggedright,singlelinecheck=off]{caption}

\makeatletter
\renewcommand{\@biblabel}[1]{\quad#1.}
\makeatother

\usepackage{lastpage,fancyhdr,graphicx}
\usepackage{epstopdf}
\fancyheadoffset[L]{2.25in}
\fancyfootoffset[L]{2.25in}
\usepackage[figuresleft]{rotating}
\usepackage{mdframed}
\usepackage{enumitem}
\usepackage{pdflscape} 

\newenvironment{compactitem}
{\begin{itemize}[nosep, leftmargin=*, topsep=0pt, partopsep=0pt, parsep=0pt, itemsep=0pt]}
{\end{itemize}}

\begin{document}
\vspace*{0.2in}

\begin{flushleft}
{\Large
\textbf\newline{Physical Models of Embryonic Epithelial Healing} 
}
\newline
\\
Rafael Franco Almada\textsuperscript{1,2,*},\text{ }
 Nuno André Miguel Araújo\textsuperscript{1,2},\text{ }
 Pedro Patrício\textsuperscript{1,3}
\\
\bigskip
\textbf{1} Centro de Física Teórica e Computacional, Faculdade de Ciências da Universidade de Lisboa, Lisboa, Portugal
\\
\textbf{2} Departamento de Física, Faculdade de Ciências da Universidade de Lisboa, Lisboa, Portugal
\\
\textbf{3} Departamento de Física, Instituto Superior de Engenharia de Lisboa, Instituto Politécnico de Lisboa, Lisboa, Portugal
\\
\bigskip

* rffalmada@fc.ul.pt

\end{flushleft}

\section*{Abstract}
Embryonic healing in epithelial tissues is distinct from adult wound healing, as it lacks inflammatory responses or immune cell recruitments, making it ideal to test models of wound healing driven primarily by epithelial dynamics. Many models have been developed to describe this process, ranging from simple mechanistic models to more elaborate multiscale simulations. We review different classes of physical models, from discrete to continuum models, and how they address key questions about the mechanics, signaling, and coordination of cells during wound closure. We highlight tensions between model complexity and interpretability and discuss recent efforts to bridge gaps across scales. Finally, we identify directions for hybrid modeling and model-experiment integration that could push forward our understanding of epithelial repair in development and disease.


\section*{Introduction}

\label{section:bio}

\begin{figure}[!ht]
    \centering
    \includegraphics[width = 0.45\textwidth]{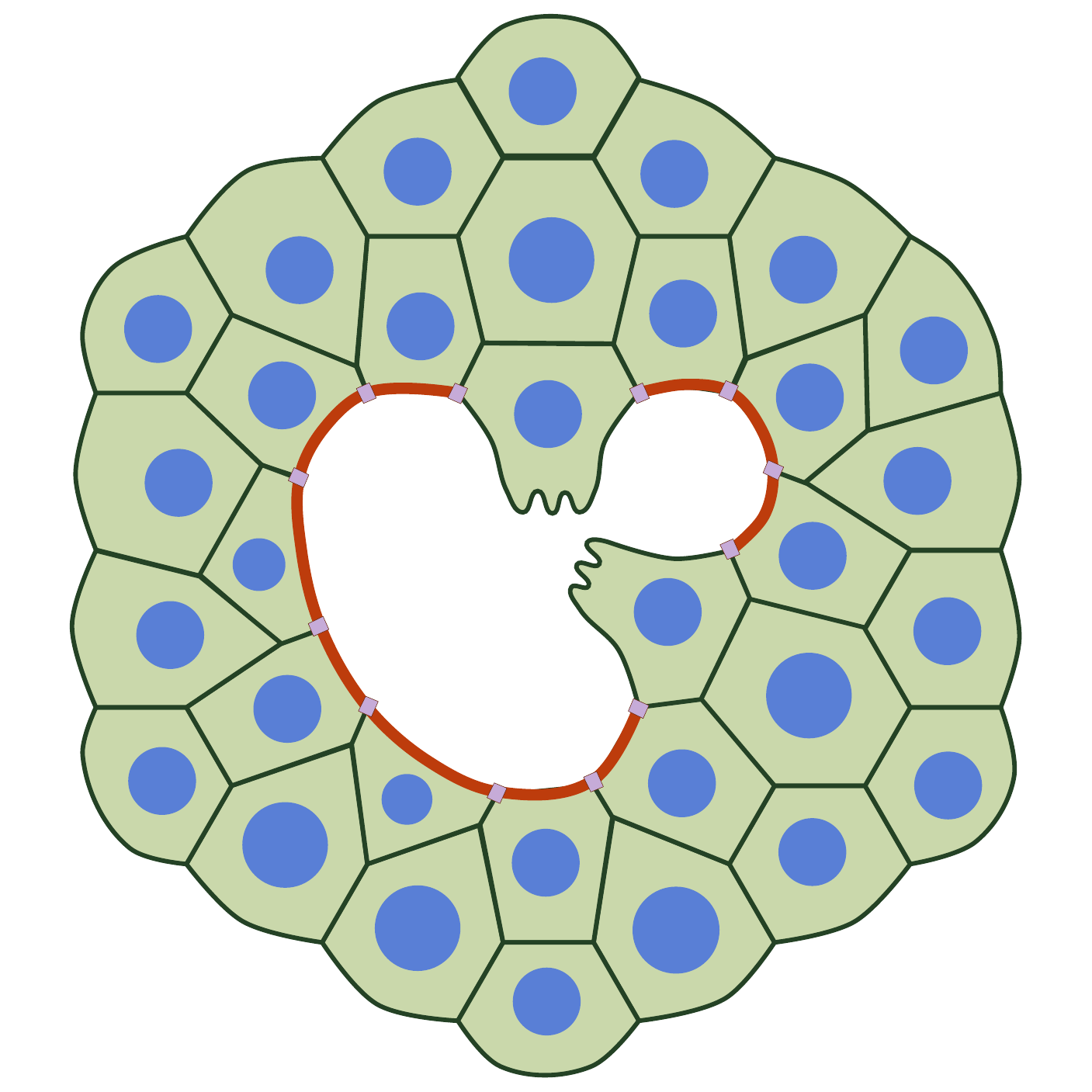}
    \caption{Wound healing in embryonic epithelial tissues, displaying both the purse-string (red cable at wound border) and cell crawling (protrusions in leading edge cells).}
    \label{fig:figure1}
\end{figure}

Embryonic wound healing in epithelial monolayers refers to the rapid repair of tissue gaps during early development, essential to preserve its integrity and barrier function.  Unlike adult wound healing, it is different in several ways: it is faster, does not leave scars, has minimized inflammatory response and has more frequent junctional rearrangements. In contrast, adult wound healing involves the recruitment of immune cells, matrix deposition by fibroblasts and scarring. With these differences, embryonic wound repair is a minimal system to study tissue repair driven primarily by epithelial dynamics. It is, therefore, ideal to test mechanistic models that aim to predict tissue closure from cellular and subcellular dynamics.
Wound closure is achieved through two primary, often coexisting, mechanisms: a purse-string around the wound edge and crawling of epithelial cells into the gap (see Fig \ref{fig:figure1}). While crawling is common to both embryonic and adult healing, the purse-string is more prevalent in embryonic healing and generally less prevalent in adult epithelia \cite{lodish, Begnaud2016, nodder1997, tamada2007two, klarlund2012dual}. 

The first mechanism, known as the purse-string, involves the formation of an actin-myosin ring around the wound gap.  Actin and myosin II accumulate rapidly at the wound edge, forming an active contractile ring that pulls the surrounding cells inward. This ring transmits force through adherens junctions, driving tissue closure \cite{Begnaud2016}. Its effectiveness depends on the local variations in cell contractility and adhesion. This mechanism is also present in cellular extrusion, where a dying (apoptotic) cell is squeezed out of the epithelium by a similar contractile ring. Dorsal closures in  morphogenesis also rely on this mechanism \cite{martin1992, danjo1998,bement1999wound, kiehart1999wound, jacinto2002, sonnemann2011, villars2022, mitchison1996, jacinto2000, wood2002, martin2004, theveneau2013}. 
The second mechanism is the collective migration of cells at the edge of the wound, generally denoted crawling \cite{poujade2007}. Cells extend protrusions called lamellipodia, adhering to the substrate and pulling themselves forward, while maintaining strong intercellular junctions through proteins such as E-cadherin and formin \cite{anon2012cell, rao2016}. During cell crawling, cells transition from the apical-basal to front-rear polarity, which requires large-scale reorganization of the cell structure and force orientation. Leading cells move into the gap, while cells further back extend lamellipodia beneath them \cite{chapnick2014}. This mechanism is effective for larger gaps, contrasting with the purse-string mechanism. Cell crawling is triggered by free space and self-polarization \cite{ravasio2015, Begnaud2016}. 

Both mechanisms depend on time-dependent cytoskeletal remodelling, adhesive dynamics and coupling between force generation and cell polarity. The purse-string requires spatially varying models of junctional tension and contractility, while crawling requires models of dynamic cell polarity and interactions with the extracellular matrix. Both are regulated by mechanical and biochemical cues, including epidermal growth factor (EGF), calcium influx (Ca\textsuperscript{2+}), reactive oxygen species (ROS) and extracellular adenosine tri-phosphate (ATP). { Laser ablation experiments have also been used to probe embryonic tissue mechanics in vivo and constrain models of epithelial wound closure} \cite{nishida1993, watanabe1987, hunter2017, carvalho2018, rothenberg2019, chen2019large, wei2020actin, jain2020, trubuil2021, rothenberg2023, arciero2011,vedula2014, vedula2015, ravasio2015regulation, haensel2018, enyedi2015,scepanovic2021, lehne2022, lee2023, brugues2014, Begnaud2016, guzman2020, ly2024, longmate2014, ma2009probing}. 

Capturing these mechanisms in models presents several challenges. To organize the modelling challenges, we group the relevant biological processes into the following domains, each associated with distinct physical mechanisms and timescales:

\begin{itemize}
    \item \textbf{Purse-string contractility and protrusive activity}: Mechanisms that involve purse-string contractility and collective motion to close the wound;
    \item \textbf{Cell mechanical behavior}: Cell motion, force generation and distribution during wound closure;
    \item \textbf{Wound geometry}: The influence of gap size and shape on the relative roles of crawling and the purse-string;
    \item \textbf{Cytoskeletal and junctional remodeling}: Regulation of cell shape, adhesion and force transmission;
    \item \textbf{Cellular migration and epithelial-to-meshenchymal transition (EMT)}: The migration of cells, including the transition between epithelial and mesenchymal states that facilitate wound closure;
    \item \textbf{The extracellular matrix (ECM) and its mechanical properties}: The ECM's role in regulating cell behavior and providing structural support, as well as how it influences tissue mechanics and cell-ECM interactions;
    \item \textbf{Cell polarity}: Spatial and dynamic organization of cells during wound healing that enables directed migration;
    \item \textbf{Chemical cues and signaling}: The signaling pathways and molecular signals that mediate the repair process, including growth factors; 
    \item \textbf{Mechanochemical feedback}: Coupling between mechanical forces and biochemical signaling pathways.
\end{itemize}

{ In addition to the purse-string and crawling mechanisms, closure of certain excisional wounds in multilayered embryonic epithelia, such as the Xenopus ectoderm, is also driven by contraction and ingression of cells exposed within the wound, forming characteristic `Y'-shapes \cite{davidson2002embryonic, hazar20153d}.}

Embryonic healing involves the coordination of physical mechanisms across coupled spatial and temporal scales. Purse-string contraction  generates edge-localized tensions over seconds to minutes, while cell crawling includes slower traction-based migration over longer ranges and timescales. Their coordination is dynamic, dependent on a wound geometry that changes over time. The lack of clear separation of scales, coupled with the active force generation and mechano-chemical feedback makes embryonic healing a classic example of a multiscale, out-of-equilibrium system. Capturing this behavior requires modeling approaches that integrate multiple levels of biological organization, from molecular scale cytoskeletal dynamics to tissue-scale stress redistribution. Physical modeling is essential not only for integrating these levels, but for identifying the mechanisms that govern closure dynamics \cite{wyczalkowski2013computational,jorgensen2016mathematical, rothenberg2019}.

This review focuses on how different models, many of which are widely used to study tissue behavior, have been applied to describe key mechanisms of embryonic healing on epithelial monolayers. A key challenge in modeling wound healing is its inherently multi-scale nature. Physical models are useful to link biological observations and measurable physical properties, which are explored in detail in later sections. We also look at how these models have helped to answer several of the key processes of embryonic healing, which challenges remain, and what might be the future direction of this growing subfield. This review does not aim to give an exhaustive overview of the different methods available to model cell migration or tissue dynamics in general, as there are already excellent reviews that cover this topic in detail, many of which are referenced throughout this review itself \cite{theveneau2013,marchetti2013, camley2017, banerjee2019continuum,alert2020}. 

{
In line with this focus, we prioritize studies where these models are applied explicitly to embryonic epithelial wound healing. For some models, wound-specific applications remain limited. In those cases, we also discuss studies using the same physical models to address closely related epithelial processes, provided that the modeled mechanisms are directly relevant to wound closure or that the original works explicitly note their applicability to healing geometries. Such inclusions are used selectively and only when necessary to represent a given modeling approach. Throughout, we restrict our discussion to epithelial monolayers, while drawing connections to morphogenetic studies only where they directly inform healing mechanisms. 

The embryonic epithelial systems covered in this review include laser-induced wounds in the \textit{Drosophila} pupal notum, epidermis and imaginal disc, with occasional consideration of other embryonic systems  such as chicks and zebrafish (\textit{Danio rerio}). For studies that do not directly address embryonic healing, the corresponding systems involve in vitro epithelial monolayers subjected to scratch assays or gap-closure geometries. These epithelial monolayers studied include, most commonly,  MDCK (Madin-Darby canine kidney)  and HaCaT (immortalized human keratinocyte) cells. These systems serve as physical models of epithelial mechanics and collective behavior relevant to wound healing, rather than as models of embryological development in the strict sense.}

In section \hyperref[section:phys]{2}, we introduce physical modeling approaches and the different ways biological tissues can be described. We will consider four categories of models: \hyperref[section:cb]{cell-based} (CB), \hyperref[section:cm]{continuum} (CM), \hyperref[section:hm]{hybrid} (HM) and \hyperref[section:data]{data-driven} (DD) models and how they have been used to study and to solve problems linked to embryonic healing. In section \hyperref[section:discussion]{3}, we compare the different models, how well they explain some aspect of embryonic healing, and we discuss what questions remain to be addressed.

\section*{Physical models}
\label{section:phys}

Describing tissue mechanics requires considering its relationship to tissue function. 
Over the past decades, numerous physical models have been developed to uncover unifying principles and highlight similarities and differences across various cellular systems \cite{camley2017,alert2020, nava2020modelling}.  
There are several large-scale dynamics at the tissue level dependent on cell type and the physical and chemical environment to which it is exposed. There are essentially two approaches to modeling tissues, depending on the chosen scale:

\begin{itemize}
    \item Cell-based models - The observed tissue behavior emerges from interactions among multiple cells, at different levels of detail. These include \textbf{lattice} models, \textbf{network} models and \textbf{phase field} models;
    
    \item Continuum models -  Tissue behavior is considered at a larger scale, using tools from \textbf{continuum mechanics} and \textbf{coarse-grained field theories}.
\end{itemize}

Neither approach is inherently superior,  but they serve different purposes depending on the research question. Cell-based models are useful for simulating collective cell migration observed in embryonic healing from individual cell interactions. Continuum models can represent large-scale deformations and mechanical properties of epithelial tissues during wound closure providing insight on the stresses and strains the tissue experiences during this process (see Table \ref{tab:table1}).  

Models typically begin with a set of assumptions that may vary in their implementation:

\begin{itemize}

    \item \textbf{Cell mechanical assumptions} - vital for accurately modeling contractile forces, as cells must deform and adhere to one another while responding to mechanical stresses:
    \begin{itemize}
        \item Individual cells are visco-elastic objects influenced by cytosketelal properties;
        \item Cells form junctions with other cells and with the substrate, there is strong intercellular adhesion and communication;
        \item Epithelial cells have a distinct polarity with specialized apical and basal surfaces. Often, models have to account for the direction-dependent mechanical interactions which result from this polarity.
    \end{itemize}
    
    \item \textbf{Topological assumptions} - essential to understand how epithelial cell cycles and junctions respond to injury and facilitate rapid wound closure:
    \begin{itemize}
        \item { Epithelial tissues form confluent monolayers - continuous sheets without holes};
         \item Cell rearrangements occur through changes in junctions and the cytoskeleton;
        \item The cell cycle involves growth, division, and apoptosis.
    \end{itemize}
    \item \textbf{Chemical assumptions} - How cells initiate and promote healing. It can also shed light on how cells communicate and coordinate their movements during embryonic healing. It involves both signaling and cell internal activity:
    \begin{itemize}
        \item Chemical signaling occurs through gap junctions and extracellular diffusion;
        \item Individual cells have metabolic activity, which may be coupled to the cell cycle.
    \end{itemize}
\end{itemize}

Implementing all these characteristics into a single model is not feasible, so simplifications and abstractions are fairly common. As the focus of our work will be on epithelial monolayers, these results may not be immediately generalizable to other tissues \cite{smallwood2009}. { Connective tissues lack the tight cell junctions which are characteristic of epithelial tissues, with their interactions being primarily mediated by the ECM, while epithelial cells have persistent junctional mechanics that couple the cells, and the ECM is interacted with at the basal face \cite{pritchard2014}}. Muscle fibers have contractile properties and spatial organization that differs significantly from epithelial sheets \cite{jaslove2018}. Neural tissues are made of neurons which have a unique morphology with axons and dendrites, with interactions mediated by synapses, rather than continuous junctions \cite{budday2020}. 

{ Considering the assumptions outlined above, we focus on multicellular embryonic wounds, which are treated physically as free boundaries within a contiguous, confluent epithelial monolayer.} In addition to the previously mentioned categories of assumptions, we may include additional properties in embryonic models. Polarity is an example of such a property, as a relevant factor in many embryonic healing processes \cite{guzman2020}. There are several polarity mechanisms that can be explored, but they can be grouped into cell-autonomous mechanisms \cite{szabo2006, basan2013, wen2021} and non-cell-autonomous mechanisms \cite{vicsek1995, rappel1999, vedel2013, nissen2018}. Choosing one mechanism constrains both the biological regimes applicable and the model's validity domain. There may be distinctions between leader cells and non-leader cells, particularly in embryonic healing processes, and different types of cells may have different responses to some signals \cite{camley2017}.

Collective motion in cells may be driven by several factors, one of them being chemical gradients, in a process described as chemotaxis. Chemical signaling during chemotaxis is usually modeled by linking extracellular chemical gradients to intracellular signaling pathways that steer cell motion.  Cells can produce their own chemoattractants and degrade them to create local gradients, resulting in cell aggregation and self-organized clusters. Meanwhile, the self-generated chemical gradients enable long-range coordination of cell movements by linking local chemoattractant production to global migration dynamics \cite{tweedy2016, proverbio2024, paspunurwar2024}. Hybrid models, commonly applied in cancer invasion studies, also provide insight into embryonic healing, bridging tissue-scale interactions, and cellular microenvironments.
 One of the challenges in modeling tissue mechanics is extending theories developed in 2D to three-dimensional contexts. This is especially relevant when scaling models from individual cells to full tissue behavior and from in vitro settings to in vivo conditions \cite{alert2020, gomezgalvez2021}. Our review mainly address results for models in 2D, however there are experimental observations which show 3D models may be more accurate to address apico-basal polarity's role during healing. { A clear example of this limitation is for cell-ECM junctions, which occur at the basal surface, and can result in out-of-plane deformations. Most of the considered models incorporate ECM through effective interactions. Additionally, effects of cells layers below are not considered, which fails to capture non-canonical closure mechanisms} \cite{davidson2002embryonic,lim2024forced}.

\begin{sidewaystable}[htbp]
    \footnotesize
    \centering

    \newgeometry{top=0.85in, bottom=0.85in, left=0.85in, right=0.85in, footskip=0.75in}
    \vspace*{2.1in}
     \begin{tabular}{|m{3em}|m{3cm}|m{4.cm}|m{1.5cm}|m{4.cm}|m{4.cm}|}\hline
     
         \textbf{Model}&\textbf{Schematic}& \textbf{Features}&  \textbf{Scale}&  \textbf{Strengths}& \textbf{Limitations}\\\hline\hline
         \textit{Lattice}& \includegraphics[width = 0.13\textwidth]{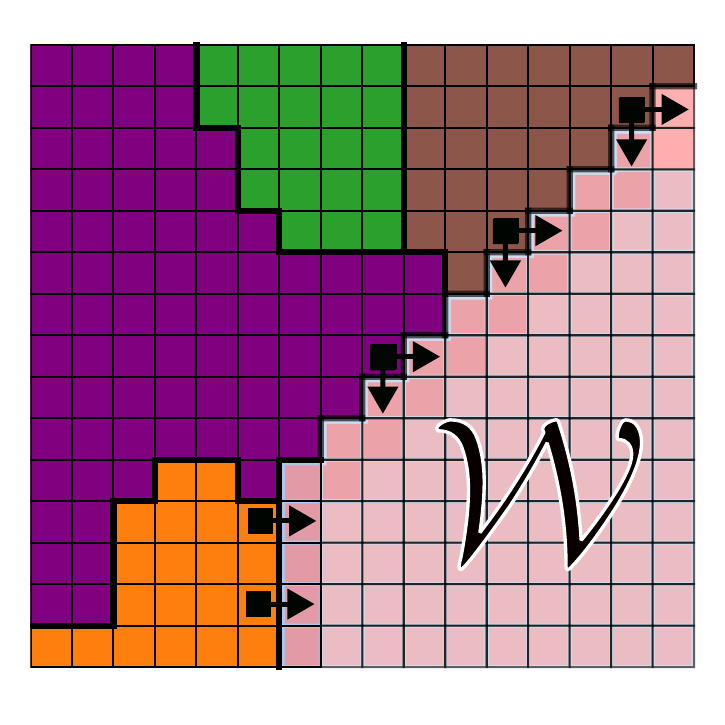} &  \begin{compactitem}
    \item Discrete lattice representation;
    \item Cells as collection of pixels;
    \item Stochastic updates via energy minimization;
    \item Cell motion via pixel rearrangements.
\end{compactitem}
&  Subcellular to cellular&  \begin{compactitem}
    \item Efficient for large cell populations;
    \item Handles cell rearrangements naturally;
    \item Can include heterogeneity and stochasticity.
\end{compactitem}& \begin{compactitem}
    \item Interactions propagate discretely;
    \item Timescale is algorithm-dependent;
    \item Discretizations limit mechanical accuracy at tissue scale.
\end{compactitem}

\\\hline
         \textit{Network}&\includegraphics[width = 0.13\textwidth]{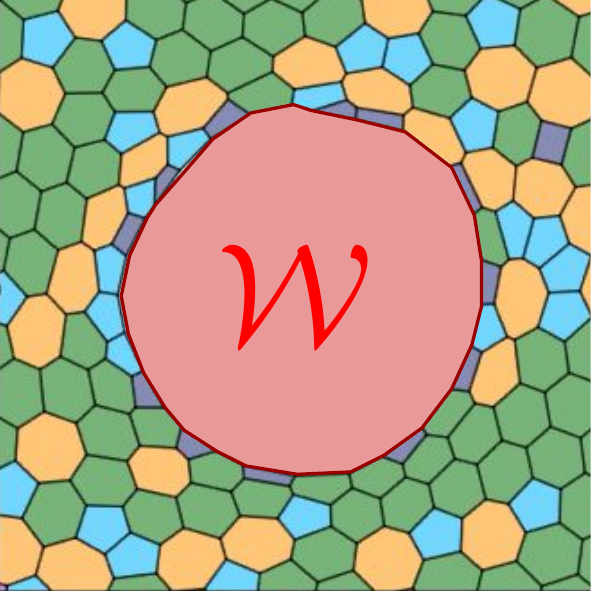}&  \begin{compactitem}
    \item Cells represented as polygons;
    \item Displacement of vertices (Vertex) via energy minimization;
    \item Displacement of centers (Voronoi) via energy minimization with geometry reconstructed;
    \item Junctional forces and contractility encoded along edges;
    \item Confluence constraint .
\end{compactitem}&  Cellular&  \begin{compactitem}
    \item Explicitly captures junctional and anisotropic tension;
    \item Handles cell rearrangements;
    \item Efficient for confluent epithelia.
\end{compactitem}
& \begin{compactitem}
    \item Coarse-grained subcellular dynamics;
    \item Cannot resolve smooth boundary deformations (protrusions);
    \item Nonconfluent tissues are not naturally represented in the models.
\end{compactitem}

\\\hline
         \textit{Phase-field}&\includegraphics[width = 0.15\textwidth]{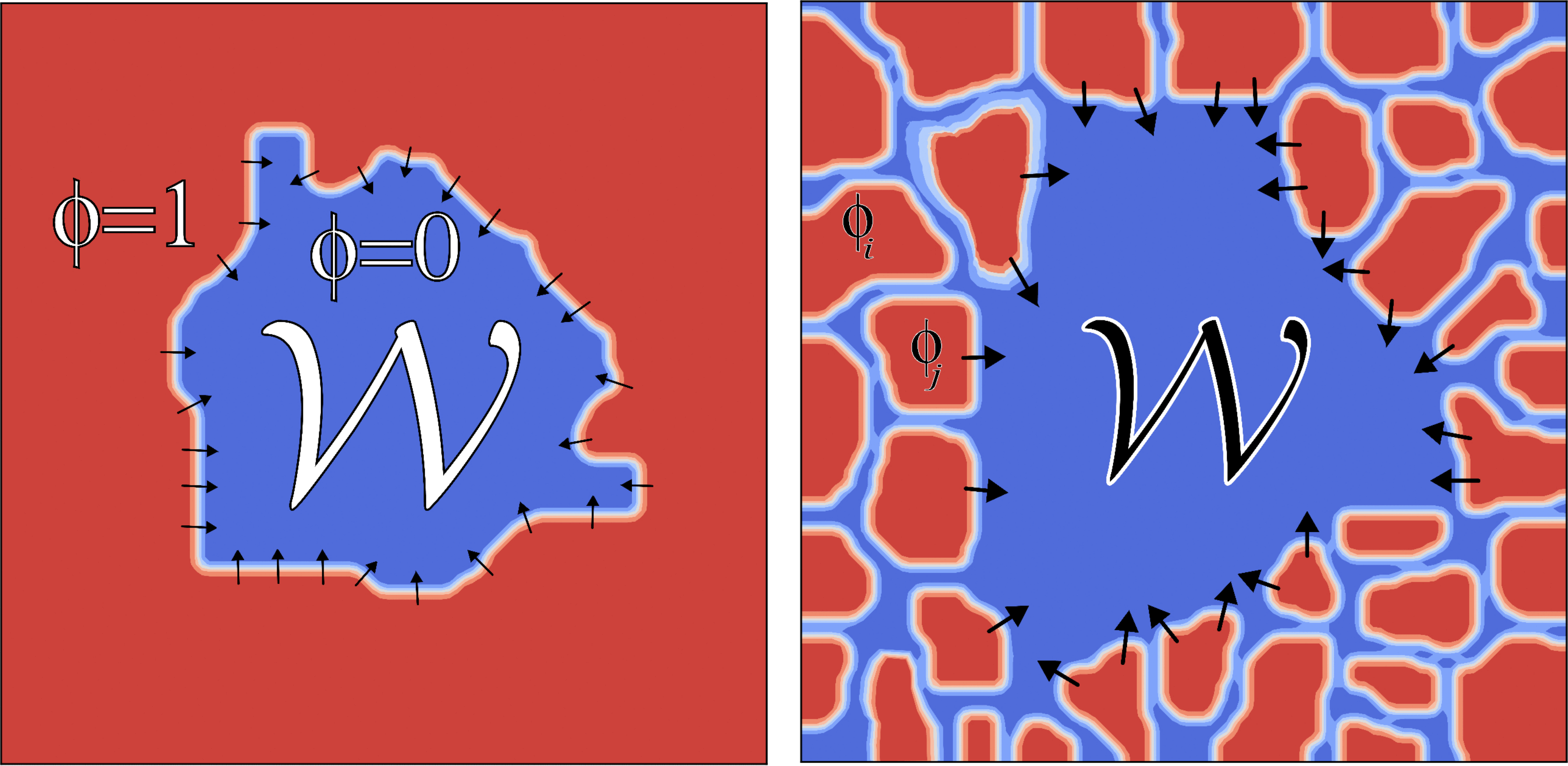} &  \textbf{Multiphase} \begin{compactitem}
    \item Cells are continuous scalar fields;
    \item Diffuse interface between cells;
    \item Cell shape and motion via field dynamics.
    
    \end{compactitem}

    \textbf{Binary}
    \begin{compactitem}
    \item Tissue and environment separation via two scalar fields;
    \item No explicit individual cells;
    \item Interface motion via Cahn-Hilliard/Allen-Cahn dynamics.
    
\end{compactitem}
   
&  Cellular (Multiphase) and Tissue (Binary) & \textbf{Multiphase} \begin{compactitem}

    \item Smooth cell boundaries;
    \item Cell deformation, growth and division emerge from continuous geometric dynamics;
    \item Can simulate many cells without explicit computational tracking.
\end{compactitem}

\textbf{Binary }

\begin{compactitem}
    \item Efficient for large-scale tissue dynamics;
    \item Captures wound interface motion and large scale instabilities;
    \item Can include active stresses.
\end{compactitem}
& 
\vspace{0.1cm}
\textbf{Multiphase}

\begin{compactitem}
    \item High computational cost for many cells;
    \item Interface thickness must be tuned carefully;
    \item Junctional mechanics are phenomenological rather than explicit.
\end{compactitem}

\textbf{Binary}
 
\begin{compactitem}
    \item Cannot resolve single-cell mechanics;
    \item Junction-level and subcellular processes coarse-grained;
    \item Not suitable to study anisotropy or heterogeneity at cellular scale.
\end{compactitem}
\vspace{0.1cm}
\\\hline
         \textit{Hydro-dynamic}&\includegraphics[width = 0.13\textwidth]{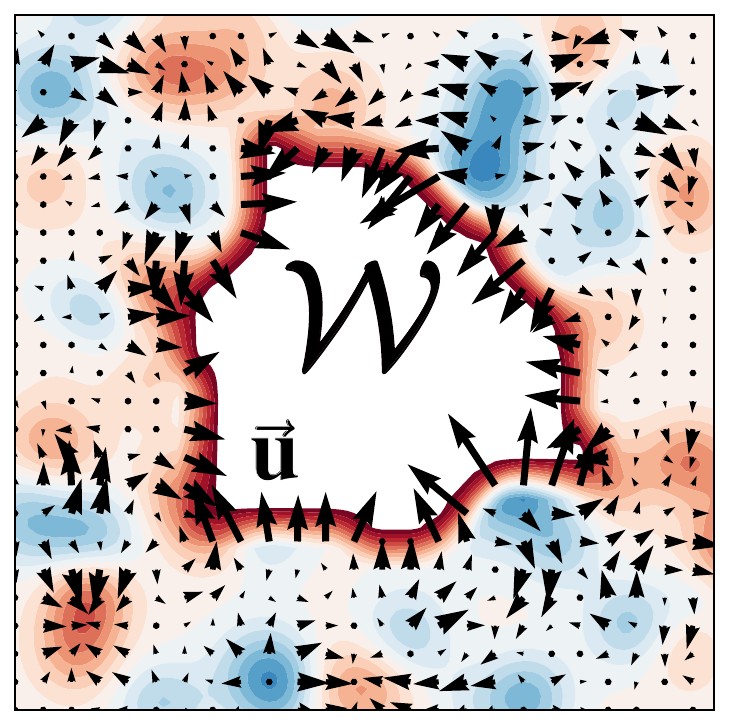}&  \begin{compactitem}
    \item Tissue-scale continuum scalar and vector fields;
    \item Flows in epithelial sheets via force-balance equations;
    \item Symmetry choice may be isotropic, polar or nematic.
\end{compactitem}
&  Tissue&  \vspace{0.2cm} \begin{compactitem}
    \item Handles large-scale collective flows and stress propagation
    \item Naturally includes active stresses
    \item Can include polarity alignment
    \item Can describe both isotropic and anisotropic tissue behaviors
\end{compactitem}
& \begin{compactitem}
    \item Subcellular details are not resolved
    \item Assumes continuum approximation
    \item Loses cellular heterogeneity
\end{compactitem}

\\\hline
         \textit{Visco-elastic}&\includegraphics[width = 0.13\textwidth]{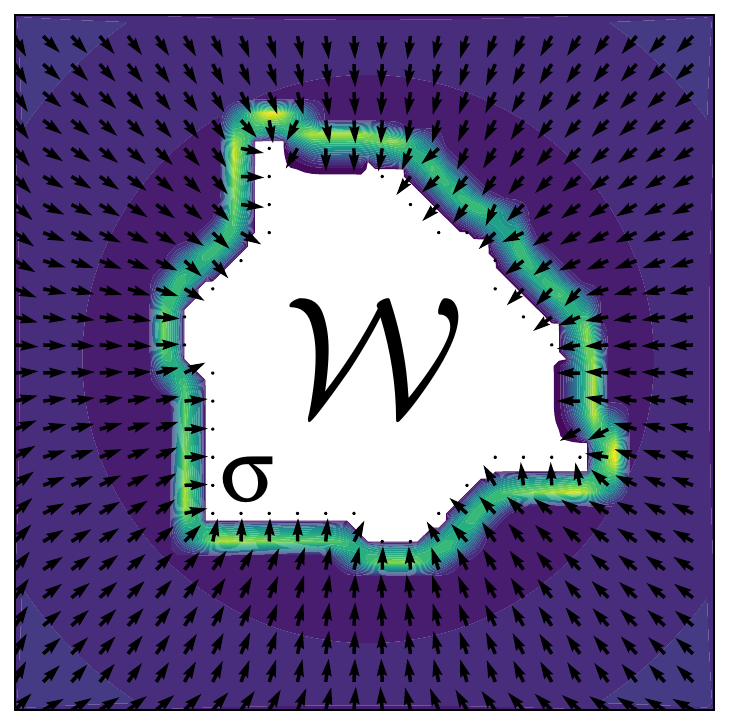}&  \begin{compactitem}
    \item Tissue treated as a viscoelastic material;
    \item Constitutive stress-strain relation;
    \item Time-dependent stress relaxation and material response;
\end{compactitem}
&  Tissue&  \vspace{0.1cm} \begin{compactitem}
    \item Simple framework to capture bulk tissue mechanics
    \item Directly relates stress and strain
    \item Easily coupled to external forces or boundary conditions
\end{compactitem}
&\vspace{0.1cm}  \begin{compactitem}
    \item Does not capture active stresses
    \item Subcellular processes are absent
    \item Does not naturally include anisotropic or orientational order
\end{compactitem}
\\\hline\hline
 
    \end{tabular}
    \vspace{0.3cm} 
    \caption{Table summarizing the main models explored in the review. The listed features correspond to the minimal structural elements of each model class; additional couplings and fields commonly used in the literature are discussed in the main text.}
    \label{tab:table1}
    \restoregeometry
\end{sidewaystable}

\newpage
\subsection*{Cell-based Models}
\label{section:cb}

In cell-based models, tissues are described as a collection of individual functional units, cells, that interact with each other. These interactions, although usually described by simple rules, may lead to complex emergent behavior that is observed at  scales much larger than the individual cell size \cite{alert2020}. The main appeal of these models comes from their conceptual simplicity and biological accuracy, where observed cellular behavior can be incorporated into the model, to see its impact at the tissue scale \cite{fletcher2022}. 

{ These models are coarse-grained at the cellular level, where internal cell dynamics are abstracted away into effective behaviors. These models assume that tissue mechanics is dominated by cell-level processes such as adhesion, cortical tension, and neighbour rearrangements, with tissue-scale behavior emerging from their collective organization.} Cellular packing also plays a central role in regulation of tissue growth, material properties and cell communication. Discrete models are very useful to understand these dynamics \cite{gomezgalvez2021}. Cell-based models are particularly useful when describing relatively small tissues, where finite-size effects and fluctuations become relevant to the observed dynamics, being particularly useful in microscopic phenomena \cite{banerjee2019continuum}.

\subsubsection*{Lattice models}

Lattice models describe tissue dynamics using a discrete spatial framework, where cells occupy specific sites on a grid, or lattice. This structure leads to localized cell interactions, leading to emergent collective behaviors that help to explain processes like cell migration and tissue organization.

\begin{figure}[h]
    \centering
    \includegraphics[width=0.45\linewidth]{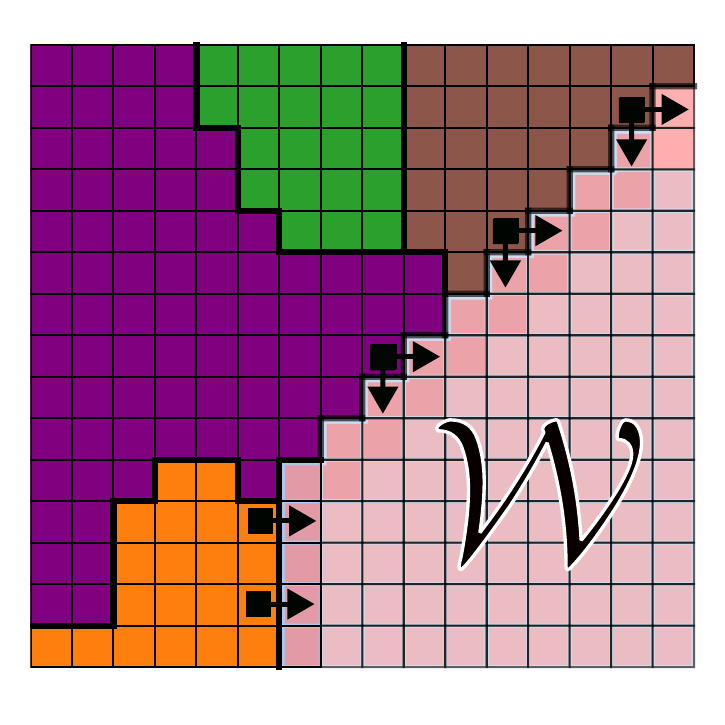}
    \caption{Schematic of a lattice model of wound healing, where $\mathcal{W}$ is the wound region and the different colored pixel regions correspond to different cells.}
    \label{fig:figure2}
\end{figure}

The \textbf{Cellular Potts Model} (CPM), also known as Glazier-Graner-{ Hogeweg} model, represents cells or subcellular components as clusters of lattice sites that can change their configuration based on energy minimization principles \cite{graner1992}. Cell rearrangement in CPM is probabilistic and modeled using the Boltzmann distribution, which reflects energy changes resulting from interactions between neighboring sites while respecting constraints on volume or area, controlled through a temperature parameter (see Fig \ref{fig:figure2}).
The system's energy $E$ is typically given by:

\begin{equation}
    E = \sum_{i,j}{J_{ij}\left(1-\delta(\sigma_i,\sigma_j)\right)} +\sum_{\sigma_i}{\lambda {\left(A(\sigma_i)-A_0(\sigma_i)\right)}^2}.
\end{equation}
The first term, based on Steinberg's \textit{differential adhesion hypothesis} \cite{steinberg1970does, foty2005, steinberg2007}, describes interactions between neighbouring lattice sites $i$ and $j$ of different domains $\sigma_i$ and $\sigma_j$, where
$J_{ij}$ determines the adhesion strength depending on cell type. The Kronecker delta function, $\delta(\sigma_i,\sigma_j)$ ensures energy contributions occur only at interfaces between different cells. The second term constrains deviations in cell area or volume, enforcing conservation through a Lagrange multiplier $\lambda$. Some models also incorporate a term representing interactions between lattice sites corresponding to intra- and extracellular components (as the extracellular matrix), usually as a modified adhesion term, with one of the domains $\sigma_i$ defined to be the cellular exterior \cite{scianna2012}.

CPM has been widely used to model cell sorting, tissue growth and cell movement within tissues, as well as morphogenesis \cite{hirashima2017, sadhukhan2021, chiang2016, durand2016, szabo2013, alsubaie2023, alsubaie2024, yamashita2024}. While lattice models are computationally efficient and allow the observation of emergent tissue behavior, it is not without drawbacks. The discretization of space may lead to artifacts in simulations, by ignoring off-lattice behaviors { and strain propagation occurs discretely, limiting mechanical accuracy at tissue-scale}. Additionally, the temperature parameter lacks a clear experimental analog, { and the implicit notion of time in the Metropolis algorithm makes dynamic interpretation difficult. While extensions such as kinetic Monte Carlo can approximate timescales, they require knowledge of event rates that are typically unknown.} And finally, it does not fully address cell-substrate interactions \cite{scianna2012, alert2020, azuaje2011}.

{Lattice models have been applied in embryonic healing and related subjects, particularly to model collective cell motion during the healing process. It has been shown that the properties of the ECM regulate the speed and direction of cell migration. Among the properties of the substrate that are key are fiber density and adhesiveness, which determines how fast the healing occurs, and fiber orientation which determines the direction of locomotion \cite{scianna2012cellular,scianna2015}. 
These models have demonstrated that an optimal balance between cell adhesiveness and contractility is necessary for wound closure to occur \cite{noppe2015}. In a particularly notable result, CPM is able to reproduce the experimentally observed universal coupling between speed and persistence, where faster moving cells change directions less frequently, a behavior which is linked to retrograde actin flow \cite{wortel2021local}. It suggests that an intermediate level of adhesiveness is required to maintain coordinated migration during embryonic healing, while allowing for flexibility in cell rearrangements \cite{roy2021intermediate}. }

\subsubsection*{Network models}

Network models are a type of off-lattice  models that represent tissues as polygonal networks and cells as polygons, where vertices correspond to junctions between three polygons, and edges define boundaries between adjacent cells. These models are effective in studying the mechanical properties of epithelial monolayers, where cell rearrangements, division and death play a key role in tissue dynamics. These different transformations typically occur through topological transitions, and can be characterized by T1 transitions (cell rearrangements), T2 transitions (cell extrusions and death), and cell divisions, which may depend on cell polarity and shape \cite{honda1978, nagai1988, cadart2014}. 

\begin{figure}[h]
    \centering
    \includegraphics[width=0.45\linewidth]{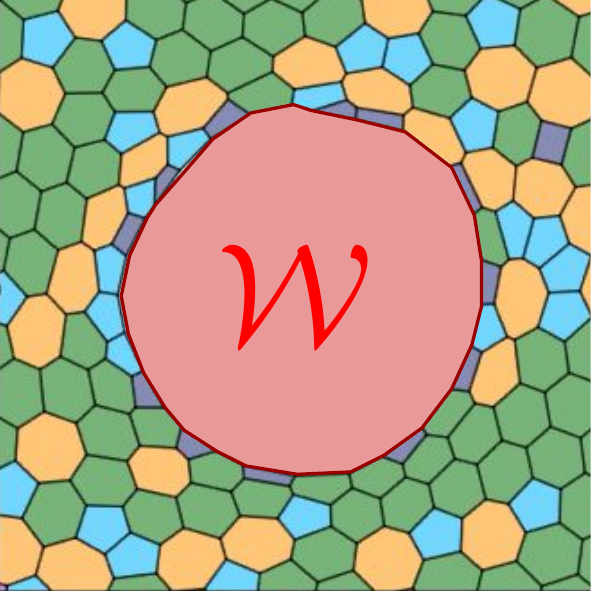}
    \caption{Schematic of a network model of wound healing, where $\mathcal{W}$ is the wound region and the different colored polygons correspond to cells with different number of neighbours.}
    \label{fig:figure3}
\end{figure}

The most widely used network model is the \textbf{vertex model}, where the tissues are typically confluent, and the vertices set the degrees of freedom of the system (see Fig \ref{fig:figure3}). The positions of these vertices evolve over time according to forces arising from mechanical interactions between neighboring cells. The energy function is typically of the form:

\begin{equation}
    E = \sum_\alpha{\left[\frac{K_\alpha}{2}{(A_\alpha-A^0_\alpha)}^2+\Gamma_\alpha {P_\alpha}^2\right]}+\sum_{\braket{i,j}}{\Lambda_{ij} L_{ij}}, 
    \label{eqn:vertexmodel2d}
\end{equation}
where, $A_\alpha$ and $P_\alpha$ are the area and perimeter of cell $\alpha$, and $L_{ij}$ is the length of a given edge $\braket{i,j}$. $K_\alpha$ are the coupling constants associated with area elasticity, $\Gamma_\alpha$ are coupling constants associated with contractility of actin ring connecting different adherens junctions and $\Lambda_{ij}$ are coupling constants associated with junctional tension in peripheral actin bundles respectively \cite{nagai2001,farhadifar2007}. The overdamped equations of motion for the vertices are:

\begin{equation}
    \frac{d \vec{\mathbf{r}}_i}{dt} = -\frac{1}{\mu}\frac{\partial E}{\partial \vec{\mathbf{r}}_i}
\end{equation}
where $\vec{\mathbf{r}}_i$ is the position of the \textit{i}-th vertex and $\mu$ is the friction coefficient. 

Cellular rearrangement in vertex models typically occur through T1 transitions, where an edge shrinks to zero and reforms in a perpendicular orientation, swapping neighbors. This model is widely used to study tissue dynamics because it accurately captures the mechanics of junctional rearrangement and the forces driving these processes. 2D vertex models can be used to explore cell packing topology in epithelial tissue and uncovering the most stable configurations \cite{staple2010, fletcher2014,cadart2014, navasedeno2017}. Vertex models have been widely applied to study a variety of biological phenomena involving cellular rearrangements, including cell sorting and tissue boundary formation \cite{landsberg2009, aliee2012, krajnc2020} and epithelial morphogenesis \cite{aegerter2010, mao2013, bi2015, erdemci2021, krajnc2021,rozman2023}. However, their explicit dependence on junction positions makes it difficult to incorporate cell motility and protrusions seen in migrating cells. Recent extensions of the vertex model, including the active vertex model implementations, have a potential use in embryonic healing applications \cite{barton2017}.

A related approach, the \textbf{Voronoi model}, defines cell boundaries through a Voronoi tessellation of cell centers instead of tracking junctions explicitly \cite{honda1978, lin2014geometrical}. Unlike vertex models, the Voronoi model have cell centers as their degrees of freedom, with cell shapes emerging from a Voronoi tessellation (hence the name of the model), and their motion governed by forces generated by neighboring cells \cite{pathmanathan2009,li2014coherent}. While this method simplifies cell rearrangements and topological transformations, it allows for a more direct integration of cell motility and polarity, making it useful for studying migration and collective behavior \cite{bi2015, manning2016, petrolli2019, lawson2022, huang2023, li2024}.

{Vertex models have been used to study how the purse-string and crawling mechanisms coordinate during closure. Simulations show that an optimal balance between these mechanisms enables an efficient closure process, robust to the mechanical properties of both the cells and the ECM \cite{staddon2018}. Neighboring cells at the wound edge have been shown to coordinate the formation of the purse-string and lamellipodia to optimize the closure rate \cite{ajeti2019}. It has been used to show that if there is a large proliferative hub during closure, a quiescent region forms, limiting division due to contact inhibition, whereas a smaller hub reduces contact inhibition’s impact \cite{zanca2022push}. While many models assume a uniform purse-string, heterogeneity in actin-myosin distribution, regulated by tension and strain, along the wound edge can improve closure \cite{zulueta2018}. Additionally, reducing junctional tension increases the intercalation rate between cells, promoting embryonic healing through an unjamming transition \cite{tetley2019}. This reduction has a greater impact on shortening closure times, as increased intercalation rates do not significantly affect stored energy but accelerates healing \cite{mosaffa2020}. It  is also possible to predict different wound outcomes, closure, shrinking, or further rupture, depending on factors such as wound size, acto-myosin contractility, and cell-cell adhesion, which are validated through experiments \cite{babu2024interplay}. Plastic deformations of cell membranes also lead to large changes in cell shapes consistent with experimental observations \cite{ton2024mechanical}. }

\subsection*{Continuum models}
\label{section:cm}

Continuum models are an alternative way to describe collective cell behavior on scales where individual cell details are less relevant than average quantities like density, velocity, or chemical concentrations and their evolution under biological and physical constraints. { Tissues are treated as  a material with effective mechanical properties like elasticity, viscosity or active stress, rather than a collection of cells}\cite{banerjee2019continuum}.

In many biological models, tissue behavior may be derived by minimizing an energy functional. Depending on the biological system, the energy functional may include terms representing cortical tension, adhesion, or curvature elasticity \cite{ehlers2011modelling, ishihara2017cells, alert2020, ackermann2021, moure2021}.  The use of a continuum model is appropriate as long as the characteristic length and geometric features of the wound are several times larger than the size of a cell \cite{banerjee2019continuum}. Therefore, discrepancies may arise as the wound approaches the closure point.

\subsubsection*{Phase field models}

\begin{figure}[!h]
    \centering
    \includegraphics[width = 0.9\textwidth]{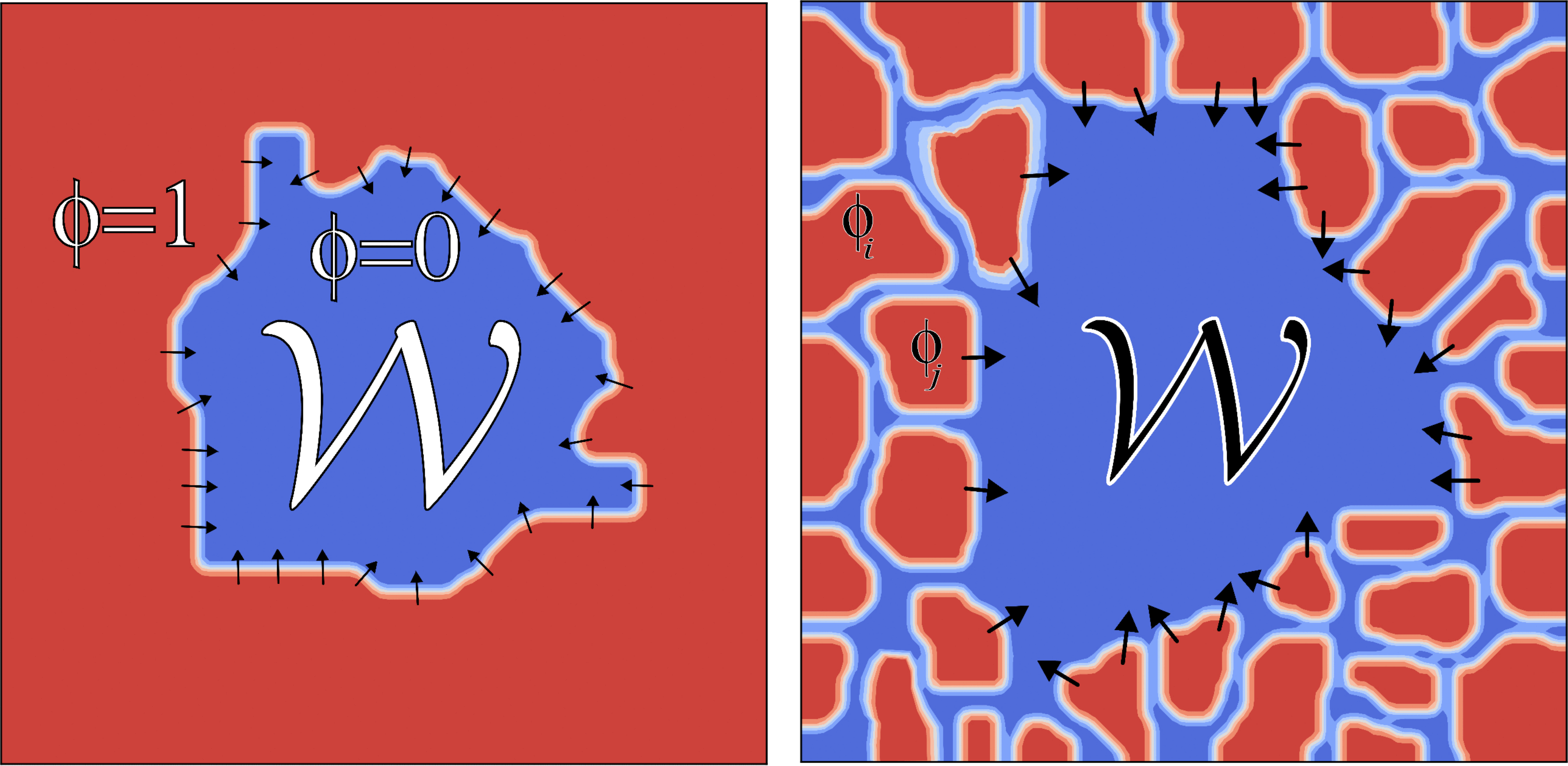} 
    \caption{(Left) Schematic of a phase field model of wound healing where $\mathcal{W}$ is the wound region where $\phi = 0$ and the tissue region is where $\phi=1$. (Right) Schematic of a multiphase field model of wound healing where $\phi_i$ is the phase field associated with cell $i$.} 
    \label{fig:figure4}
\end{figure}

We start by looking at phase field models as they are at the interface between cell-based and continuum tissue models. Phase field models are widely used to simulate dynamic interfaces in tissues because they represent boundaries implicitly, removing the need to track their positions explicitly (see Fig \ref{fig:figure4}). In these models, a phase field or order parameter $\phi$ transitions smoothly from 1 inside the tissue to 0 outside, allowing for a continuous representation of tissue structure \cite{camley2017}. 

The evolution of the phase field is governed by a free energy functional, as seen in \textbf{Cahn-Hilliard} or\textbf{ Allen-Cahn} formulations \cite{cahnhilliard1958, allencahn1975}. The free energy functional is commonly given by:  

\begin{equation}
    \mathcal{F}_\text{L} = \int_\Omega d^2\vec{\mathbf{r}} \left[ \frac{\epsilon}{2} (\vec{\nabla} \phi)^2+ f(\phi)\right].
\end{equation}
Here, the first term accounts for interfacial energy, while the second term incorporates cell-cell adhesion, surface tension, and tissue elasticity, often modelled by a Landau potential energy $f(\phi) = a(1-\phi^2)^2$, where $a$ is a proportionality constant, ensuring $f$ has units of energy density. The parameter $\epsilon$ controls the interface width, smoothing sharp transitions \cite{moure2021phase, meissner2024introduction}.  In the Allen-Cahn formulation (based on the Ginzburg-Landau equation), or dissipative dynamics, the phase field evolves according to:

\begin{equation}
    \frac{\partial \phi}{\partial t} = -\kappa\frac{\delta\mathcal{F}_\text{L}}{\delta \phi},
\end{equation}
where the left side describes the change of the phase field over time and the right-hand operator $\frac{\delta}{\delta \varphi}$ is the variational derivative of the free energy $\mathcal{F}_\text{L}$, defined on the tissue domain $\Omega$, with $\kappa$ being a constant that describe the timescale of the change in $\phi$. In the Cahn-Hilliard formulation, or conservative dynamics, the evolution follows:

\begin{equation}
    \frac{\partial \phi}{\partial t} = D \vec{\nabla}^2\left(\frac{\delta\mathcal{F}_\text{L}}{\delta \phi}\right),
\end{equation}where $D$ in this case represents a diffusion coefficient. While both formulations can be applied to embryonic healing, the choice between Allen-Cahn and Cahn-Hilliard depends on the biological assumptions of the model. The Allen-Cahn formulation is typically suited to cases where local interface dynamics dominate and mass exchange may occur, while the Cahn-Hilliard formulation is preferable when conserving cell density or volume is essential, as in confluent tissues during wound closure \cite{moure2021phase}.

A more detailed version of this framework is the \textbf{multiphase field} approach, which models tissues as collections of interacting phases, each representing an individual cell. { This is effectively a cell-based model.} Instead of modeling smooth boundary dynamics, they explicitly describe interactions between phases, using conservation laws and enforcing total area and/or volume conservation through additional constraints. A common formulation for the free energy in this case is:

\begin{equation}
    \begin{split}
    \mathcal{F} &= \sum_{i} \left[ \mathcal{F}_\text{L}(\phi_i)+ \mathcal{F}_\text{area}(\phi_i) + \mathcal{F}_\text{cell-cell}(\phi_i)\right]  \\ \mathcal{F}_\text{cell-cell}(\phi_i) &= \sum_{j\neq i}\left\{ \int{d^2\vec{\mathbf{r}}\left[\kappa_i \phi^2_i \phi^2_j - \tau_a|\vec{\nabla}\phi_i|^2 |\vec{\nabla}\phi_j|^2\right]}\right\}.
    \label{eqmultiphase1}
    \end{split}
\end{equation}
In this formulation, in addition to the Landau free energy, there is an area conservation term, defined as:
\begin{equation}
    \mathcal{F}_\text{area}(\phi_i) = \mu_\mathcal{A} \left(1-\frac{1}{\pi R^2}\int_\mathcal{A}\phi_i^2 d^2\vec{\mathbf{r}}\right) 
\end{equation}
which penalizes deviations from the preferred area $\pi R^2$, with a modulus $\mu_\mathcal{A}$, and an explicit cell-cell interaction term, given by the second line of Eq. (\ref{eqmultiphase1}). The first interaction term is repulsive, with coupling strength $\kappa_i$, while the second is adhesive, with coupling strength $\tau_a$. In biological terms, $\kappa_i$ may represent contact inhibition forces, while $\tau_a$ corresponds to cellular adhesion \cite{moure2021phase, alert2020}.  Here the dynamics of cell shapes are governed by the following expression:

\begin{equation}
    \frac{\partial \phi_i}{\partial t} + \vec{\mathbf{v}}_i\cdot \vec{\nabla}\phi_i = -\frac{\delta \mathcal{F}}{\delta \phi_i}
\end{equation}
The left side of the equation describes the change of the phase field due to its evolution over time and transport by a velocity $\vec{\mathbf{v}}_i$ \cite{landau2013fluid, giaquinta2004calculus}. $\vec{\mathbf{v}}_i$ may be given by the interaction force $ \vec{\mathbf{F}}_i^\text{int}/\xi$ with a friction coefficient $\xi$. The interaction force is given by:
\begin{equation}
    \vec{\mathbf{F}}_i^\text{int} = \int_{\Omega_i}{\frac{\delta \mathcal{F}_\text{cell-cell}}{\delta \phi_i} \vec{\mathbf{\nabla}}^2\phi_i d^2\vec{\mathbf{r}}},
\end{equation}
and describes the interactions due to boundary overlaps. $\vec{\mathbf{v}}_i$ may also be given by active motion of the cells, with a rule of $v_0 \hat{\mathbf{p}}_i$ where $v_0$ is a propulsion constant and $\hat{\mathbf{p}}_i$ is a polarity vector \cite{camley2017}. 
These models have been used to simulate a variety of processes in epithelial tissues, including cell sorting, migration, tissue organization and collective mechanical interactions  \cite{nonomura2012study, lober2015collisions, wenzel2019topological, loewe2020solid, zhang2020active, perez2024deposited, graham2024cell}. Unlike network or lattice-based models, multiphase field models can naturally capture complex topological changes like cell fusion, splitting or boundary merging. It is computationally more expensive than network models, as it requires solving PDEs and very fine spatial resolution. Despite this drawback, multiphase approaches effectively describe nonconfluent tissues, being useful for modeling epithelial-to-mesenchymal transitions (EMT) \cite{moure2021phase, chiang2024multiphase}.

{Phase fields have been relatively underexplored in embryonic healing studies of embryonic epithelia, but recent work has given some relevant insights.  For instance, elastic mismatch between neighboring cells has been shown to enhance the collective mobility of the monolayer \cite{palmieri2015multiple}. Simulations can also capture leadership dynamics observed experimentally and suggest that tissue surface tension depends on the balance between adhesion and cortical tension \cite{najem2016phase}. A major challenge in embryonic healing is understanding how cells choose between purse contraction and crawling for closure. Curvature sensing has been proposed as a mechanism for this decision-making process, though whether it operates at a single-cell or multicellular level remains unclear. Preliminary results in phase field simulations suggest that multicellular curvature sensing is efficient in coordinating the different mechanisms \cite{feng2023physical}. Additionally, recent results have explored how wound geometry and curvature influence closure dynamics, further refining our understanding of the process \cite{pozzi2024geometric}. }

\subsubsection*{Hydrodynamic models}

\begin{figure}[h]
    \centering
    \includegraphics[width=0.45\linewidth]{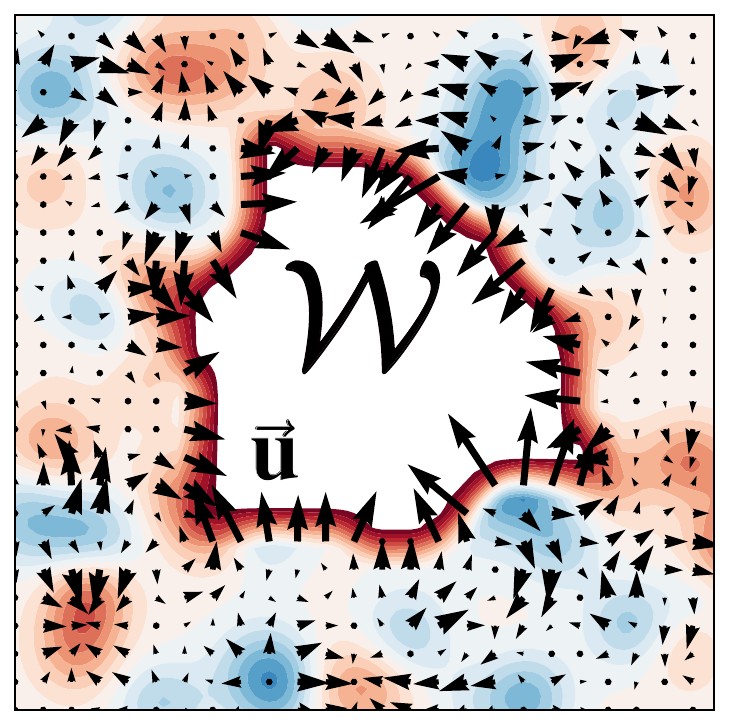}
    \caption{Schematic of an active hydrodynamic model of wound healing. $\mathcal{W}$ corresponds to the wound region, and $\vec{\mathbf{u}}$ is the velocity field, driven by active stress (blue and red regions).}
    \label{fig:figure5}
\end{figure}

In hydrodynamic tissue models, cells are no longer individualized. Instead, they describe tissues as viscous or viscoelastic fluids, giving a framework to study tissue flow, stress distribution and large-scale deformations using principles from classical hydrodynamics and associated computational tools (see Fig \ref{fig:figure5}). Unlike passive fluids, tissues exhibit active stresses driven by cell motility and contractile forces, which must be incorporated into the hydrodynamic equations \cite{marchetti2013, banerjee2019continuum}.

A fundamental equation in this approach is the continuity equation for cell density, which accounts for diffusion, cell proliferation and apoptosis:

\begin{equation}
    \frac{\partial \rho }{\partial t} + \vec{\nabla} \cdot(\rho \vec{\mathbf{v}}) = (k_p-k_a)\rho
\end{equation} 
where $\rho(\vec{\mathbf{r}},t)$ is the tissue density field, $\vec{\mathbf{v}}(\vec{\mathbf{r}},t)$ is the tissue velocity field and $k_p$ and $k_a$ denote the proliferation and apoptosis rates, respectively.
$\vec{\mathbf{v}}$ evolves according to modified Navier-Stokes equations that incorporate active stresses, alongside pressure and viscosity terms of classical hydrodynamics \cite{liverpool2006rheology, joanny2007hydrodynamic}: 

\begin{equation}
    \rho\left(\frac{\partial}{\partial t} + \vec{\mathbf{v}} \cdot \vec{\nabla}\right) \vec{\mathbf{v}} = -\vec{\nabla} p + \eta \vec{\nabla}^2 \vec{\mathbf{v}} + \vec{\nabla} \cdot \sigma^\text{A} .
\end{equation}
The left side of the equation corresponds to the material or convective derivative, while the right side includes the influence of the pressure field $p(\vec{\mathbf{r}},t)$ and of viscosity (coupled to a dynamic viscosity coefficient $\eta$). $\sigma^\text{A}$ is a stress tensor typically dependent on the local orientation or polarity of cells, reflecting internal contractile forces such as actomyosin activity (e.g. $\sigma^\text{A} = \zeta \mathbf{Q}$, where $\zeta$ is an activity coefficient that controls the magnitude of active stress, and  $\mathbf{Q}$ is the nematic tensor, which we will discuss further below). In biological tissues, inertial effects are typically negligible due to low Reynolds numbers, allowing for simplification to overdamped or Stokes-flow regimes, where $\vec{\mathbf{v}}\cdot \vec{\nabla}\vec{\mathbf{v}}$ can be ignored, in many practical applications \cite{alert2022active}.

This makes hydrodynamic models particularly useful in wound closure studies, where large-scale collective flows dominate. Below we discuss two particular classes of hydrodynamic models that are relevant to modelling tissue dynamics: active polar and nematic models.

\textbf{Polar models} represent cells or agents with a specific orientation in space, but with a focus on polarity rather than nematic alignment. In these models, the cells are typically represented as having a defined directionality, governed by an orientation field $\vec{\mathbf{p}}$. The polarity is important in systems where cells exhibit directed motion or collective polarization without the need for complex nematic order \cite{marchetti2013, camley2017, alert2020}.

These models often use a continuity equation to describe the dynamics of the polar vector field, which evolves due to local alignment interactions, external signals, or external forces, as:

\begin{equation}
    \left(\frac{\partial}{\partial t} + \vec{\mathbf{v}}\cdot\vec{\nabla}\right) \vec{\mathbf{p}}+\mathbf{\Omega} \cdot \vec{\mathbf{p}}= - \frac{\delta \mathcal{F}}{\delta \vec{\mathbf{p}}} - \nu_p  \,\mathbf{E}\cdot \vec{\mathbf{p}},
\end{equation}
The left side of the equation denotes the corotational derivative, which consists of the material derivative with an additional term coupling the polarity with the vorticity tensor, $\mathbf{\Omega} =  (\vec{\nabla} \vec{\mathbf{v}} - \left(\vec{\nabla} \vec{\mathbf{v}})^T\right)/2$. In the right side, the first term is given by the functional derivative of the free energy, given by the Frank free energy expansion in $\vec{\mathbf{p}}$.  The second term, is a coupling between the polarity field and the flow, represented by the strain-rate tensor $\mathbf{E} =  (\vec{\nabla} \vec{\mathbf{v}} + \left(\vec{\nabla} \vec{\mathbf{v}})^T\right)/2$; here, $\nu_p$ is the flow coupling constant \cite{marchetti2013, degennes1993}. The free energy is given by an expression of the following form:
\small
\begin{equation}
    \begin{split}
        \mathcal{F} = \int\left(\frac{K_1}{2}(\vec{\nabla} \cdot \vec{\mathbf{p}})^2 + \frac{K_2}{2}[\vec{\mathbf{p}}\cdot (\vec{\nabla} \times \vec{\mathbf{p}})]^2 \right.  \left. + \frac{K_3}{2}[\vec{\mathbf{p}}\times (\vec{\nabla} \times \vec{\mathbf{p}})]^2  - \frac{1}{2}h_{||}^0\vec{\mathbf{p}}^2 \right)d^2\vec{\mathbf{r}}
    \end{split}
\end{equation}
\normalsize
The first three terms are the free energy associated with splay, twist and bend deformations, the fourth term ensures that the polarization is a unit vector. 

Polar models are frequently applied to simpler tissue structures or situations where orientation is critical, but long-range order (like that seen in nematics) is not necessary.

\textbf{Active nematic models}, explicitly incorporate cell orientation and alignment in addition to tissue flow. Cells are treated as elongated agents interacting through cell-cell junctions, with their collective orientation represented by a director field $\hat{n}$ \cite{xi2019material, saw2018}. The nematic tensor is defined as $\mathbf{Q} = S(\mathbf{\hat{n}}\mathbf{\hat{n}} - 1/d \mathbf{I})$, where $S$ is the scalar order parameter, $\mathbf{\hat{n}}$ is the director field, and $d$ is the spatial dimension. This tensor captures both the magnitude and direction of local alignment. Active stresses in nematic tissues arise due to local misalignment, which increases the free energy and drives mechanical responses. The active stress is incorporated in the Navier-Stokes equation \cite{balasubramaniam2022, doostmohammadi2018active}.

The nematic tensor evolves according to the Beris-Edwards equation, which accounts for both advection and local relaxation \cite{edwards1990, thampi2016}:

\begin{equation}
    \left(\frac{\partial}{\partial t} + \vec{\mathbf{v}}\cdot\vec{\nabla}\right)\mathbf{Q} - \mathbf{S} = \Gamma \mathbf{H}.
\end{equation}
Here, $\mathbf{S}$ is the co-rotation term and describes the effect of velocity gradients on the orientation, described by the following equation:
\small
\begin{equation}
    \begin{split}
        \mathbf{S} = (\lambda \mathbf{E}+\mathbf{\Omega})\left(\mathbf{Q}+\frac{1}{d}\mathbf{I}\right) + \left(\mathbf{Q}+\frac{1}{d}\mathbf{I}\right) (\lambda \mathbf{E}+\mathbf{\Omega})-2\lambda\left(\mathbf{Q}+\frac{1}{d}\mathbf{I}\right) (\mathbf{Q}:\vec{\nabla}\vec{\mathbf{v}}) 
    \end{split}
\end{equation}
\normalsize
where $\lambda$ is the alignment parameter which controls the coupling between the nematic tensor and the velocity gradients. $\mathbf{H} = - \frac{\delta \mathcal{F}}{\delta \mathbf{Q}} + \frac{1}{3}\mathrm{Tr}(\frac{\delta \mathcal{F}}{\delta \mathbf{Q}})\mathbf{I} $ is the molecular field that ensures that the tissue relaxes to minimize the free energy $\mathcal{F}$, associated with a rotational viscosity $\Gamma$. 
The free energy itself has two components:
\small
\begin{equation}
    \mathcal{F} = \int\left(\frac{A}{2}\text{tr}(\mathbf{Q}^2) + \frac{B}{3}\text{tr}(\mathbf{Q}^3)+\frac{C}{4}(\text{tr}(\mathbf{Q}^2))^2 + \frac{K}{2}(\vec{\nabla}\mathbf{Q})^2\right) d^2\vec{\mathbf{r}}
\end{equation}
\normalsize
The first 3 terms, the bulk free energy, govern the degree of local alignment and is described as a Landau-de Gennes form, in products of $\mathbf{Q}$ with scalar symmetry, where $A$, $B$ and $C$ are phenomenological coefficients \cite{degennes1993}. Spatial variations in orientation are penalized through the fourth term, which accounts for splay (divergence), bend (curvature), and twist distortions in the nematic field, with an elastic constant $K$ \cite{frank1958}.

{Hydrodynamic, polar and nematic models have been extensively applied to understand the mechanisms of cell migration { although not much use has been directed towards embryonic healing}. Early hydrodynamic models demonstrated how monolayer elasticity, adhesive tension, and lamellipodial crawling interact during closure, revealing that partial closure can occur even without lamellipodia \cite{arciero2011}. Further studies highlighted the role of gap shape and area in determining closure speed, with crawling aiding closure while adhesion hinders it \cite{arciero2013using}. More recent hydrodynamic frameworks have studied the influence of extracellular signal-regulated kinase (ERK) concentration fields on cell migration \cite{asakura2021hierarchical}. A continuum model linking single-cell behavior to collective migration showed that  the activity of PIEZO1, a mechanosensitive ion channel, inhibits leader cell formation and suppresses directionality, emphasizing its inhibitory role in reepithelialization \cite{chen2024piezo1}.

Active polar and nematic models have shown how active stress governs cell polarization patterns \cite{he2020theoretical}. These models also demonstrated that gap size, active contractility, and purse-string contraction influence the propagation of the wound front, with active contractility potentially leading to fingering instabilities at the edge \cite{zhao2023analytical, berlyand2024bifurcation}.
At the subcellular level, nematic models have been used to study how the actin cytoskeleton behaves as a nematic fluid, influencing migration modes. For instance, curvature-induced tension anisotropies generate different actin flow patterns in concave vs. convex wound geometries, affecting how cells collectively polarize and migrate \cite{chen2019}.
}

\subsubsection*{Elastic and viscoelastic models}

\begin{figure}[!h]
    \centering
    \includegraphics[width = 0.65\textwidth]{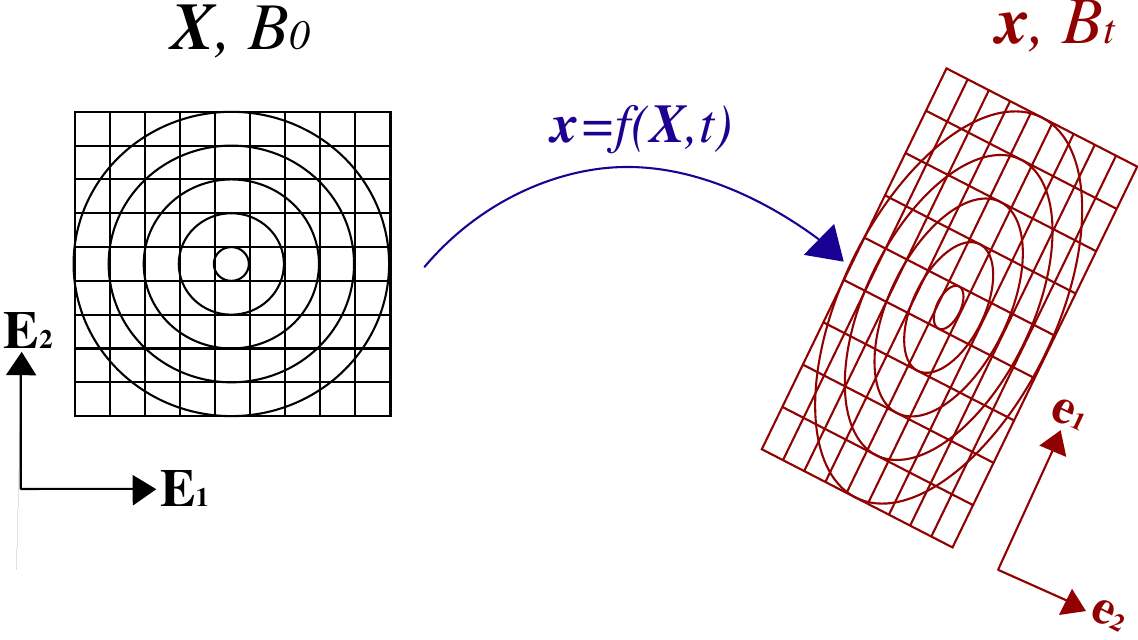}
    \caption{2D deformation of a continuous body from an undeformed configuration $B_0$ to a deformed configuration $B_t$. $\textbf{\textit{X}}$ and $\textbf{\textit{x}}$ are the positions of an arbitrary point in the body in the undeformed and deformed configurations, respectively. In the example above the body has experienced a pure shear and a rotation.}
    \label{fig:figure6}
\end{figure}

In continuum mechanics, we distinguish between material or Lagrangian coordinates $\mathbf{X} = X_I \mathbf{E}_I$, which label each material point in the initial undeformed configuration $B_0$, and spatial or Eulerian coordinates $\mathbf{x} = x_i e_i$, which label the same point at the current deformed configuration $B_t$. The two coordinates are related to each other through a mapping $\mathbf{x} = f(\mathbf{X},t)$, which can be linearized to describe stretching and rotations through the deformation tensor $\mathbf{F} = \frac{\partial \mathbf{x}}{\partial \mathbf{X}}$ (see Fig \ref{fig:figure6}).  In this way, we are able to define constitutive laws in reference to either the original or deformed configuration, as seen in the following sections.

If we describe the tissue as a deformable material, we can use constitutive relations to describe how tissues respond to different types of mechanical stress \cite{machado2015emergent, luciano2022appreciating, mierke2021bidirectional}.
Tissues are generally modeled using elasticity or viscoelasticity, each suited to specific tissue behaviors based on immediate or time-dependent responses under deformation. 
\textbf{Elastic models} are best suited for tissues that rapidly return to their original shape. An isotropic elastic material is classically described by Hooke's law:

\begin{equation}
    \sigma = \frac{E}{1+\nu}{\varepsilon} + \frac{E \nu}{(1+\nu)(1-2 \nu)}\mathrm{tr}({\varepsilon})\mathbf{I}
\end{equation}
where $\varepsilon$ is the infinitesimal strain tensor, $E$ is Young's modulus, $\nu$ is the Poisson ratio \cite{Slaughter2002}. While linear elasticity may be used in some scenarios, tissues commonly have a nonlinear elastic response, which can be better modelled with hyperelasticity models that account for large deformations using nonlinear strain energy functions \cite{haas2019nonlinear, kida2018continuum, barry2022continuum}. A common hyperelastic model is the neo-Hookean model, given by the strain energy density $w$:

\begin{equation}
    w = \mu_1(J^{\frac{2}{3}}I_1-3) + \frac{\mu_2}{2}(J-1),
\end{equation}
where $I_1 = \text{tr}(\mathbf{B})$ is the first invariant of the left Cauchy-Green strain tensor $\mathbf{B} = \mathbf{F}\mathbf{F}^\text{T}$, $J = \det(\mathbf{F})$ is the determinant of $\mathbf{F}$ where $\mathbf{F}$ is the deformation tensor, and $\mu_1$, $\mu_2$ are material constants. The Cauchy stress is given by:

\begin{equation}
    \sigma = \frac{1}{J} \frac{\partial w}{\partial \mathbf{F}}\cdot \mathbf{F}^\text{T}.  
\end{equation}
This model can be shown to reduce to the classical Hooke's law for small deformations \cite{rivlin1948, khaniki2022}.

\begin{figure}[h]
    \centering
    \includegraphics[width=0.4\linewidth]{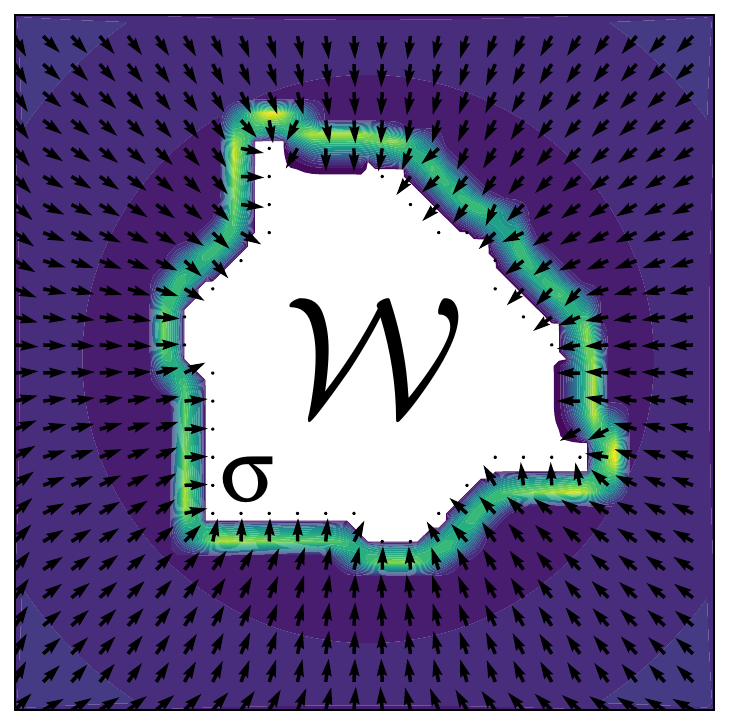}
    \caption{Schematic of a viscoelastic model of wound healing. Where $\mathcal{W}$ is the wound region and displacement is driven by the internal tissue stress $\sigma$.}
    \label{fig:figure7}
\end{figure}

\textbf{Viscoelastic models} capture both short-term elastic recovery and long-term viscous flow, closely mirroring how biological tissues spread and recover shape under sustained forces (see Fig \ref{fig:figure7}).

The Kelvin-Voigt and Maxwell models are commonly used formulations, each balancing elastic and viscous contributions depending on the dominant mechanical behavior in a given healing scenario. The viscoelastic nature of these models makes them useful to model collective cell migration, where cells adhere to each other and substrate interactions induce both rapid and sustained deformations \cite{alert2020, pajic2023physics, tlili2020migrating}.

We will use the scalar formulation of viscoelastic models for illustrative purposes, therefore the $\sigma$ and $\varepsilon$ in Eqs. (\ref{eqn:maxwellvisco}) and (\ref{eqn:kelvinvoigt}) are the scalar stress and strain respectively. The Maxwell model \cite{banks2011brief} assumes a damper and spring arranged in series, and the resulting expression describes the evolution of stress:

\begin{equation}
    E \frac{\partial \varepsilon}{\partial t} = \frac{\partial \sigma}{\partial t} + \frac{E}{\eta} \sigma  ,
    \label{eqn:maxwellvisco}
\end{equation}
where $\eta$ is the viscosity, and $E$ is the elastic modulus of the material. In the Kelvin-Voigt model \cite{dill2006continuum, carvalho2021}, the damper and the spring are in parallel to each other, and the resulting expression is:

\begin{equation}
    \eta\frac{\partial \varepsilon}{\partial t} = \sigma-E \varepsilon .
    \label{eqn:kelvinvoigt}
\end{equation}
These are the simplest viscoelastic models, though more complex generalizations are often employed to describe tissues \cite{azeyanagi2009, tsukune2015automated, du2023modeling,pajic2024friction}. Given that the tissue is a continuous material, it obeys the continuity equation \cite{lautrup}. As with hydrodynamic models, inertial terms can be neglected, which lead to the Cauchy's momentum equation:
\begin{equation}
    \vec{\nabla} \cdot \sigma + \rho \vec{\mathbf{f}} = 0.
\end{equation}
In 2D, tissue boundaries are explicitly represented as contours or curves, enabling direct control over interface evolution. The evolution of the contour, which is often determined by energy minimization or the balance of local forces at the wound edge, gives a detailed look of wound-edge dynamics. However they are often coupled to the dynamics of the bulk of the tissue. Typically the boundary evolution is given by:

\begin{equation}
    \frac{\partial \vec{\mathbf{x}}_\text{C}}{\partial t} = \left(\gamma \kappa + \lambda \right) \mathbf{\hat{n}}+\sigma\cdot\mathbf{\hat{n}}
\end{equation}
where $\vec{\mathbf{x}}_\text{C}$ is the contour position, $\gamma$ represents a surface tension, $\kappa$ is the local signed curvature, $\lambda$ may represent forcing terms and $\sigma$ represents internal cellular stresses acting on the boundary \cite{hutson2003forces, almeida2011mathematical}. These equations can model tissue dynamics driven by cell proliferation, migration, and reorganization along the wound boundary.

{Continuum mechanics models have long been used to describe various aspects of embryonic development and embryonic healing. One of the earliest physical models of wound closure described how stress-induced microfilament alignment leads to actin-myosin cable assembly \cite{sherratt1992mathematical}. This approach successfully captures complex behaviors, including biphasic contraction and transient opening at low tissue tension \cite{wyczalkowski2013computational}. A continuum model has been used to describe experimentally observed traction forces at the wound edge, to describe two observed migration modes associated with cell crawling and purse-string contraction \cite{brugues2014}. In non-adherent substrates, where crawling is not possible, purse-string contraction is the primary mechanism for closure. A continuum mechanics model predicts a reinforcement mechanism in which large-scale remodeling of the actin purse string promotes closure, with the initiation of healing depending on wound size \cite{vedula2015, wei2020actin}.
Beyond purse-string dynamics, continuum models have shown that tissue mechanics alone, independent of external signaling, can regulate cell motion during crawling \cite{lee2011crawling}. These models also establish how wound geometry and border curvature regulate the coordination between the different closure mechanisms \cite{ravasio2015}. The evolution of wound shape under these forces approaches a circular form, with closure time following predictable scaling laws \cite{he2019curvature}.
Tissue friction has been identified as a key factor in regulating closure dynamics, influencing both protrusive strength, early tissue displacement and wound recoil relative to viscosity \cite{cochet2014,carvalho2021}. Extensions to three-dimensional models reveal that stress distribution during wound closure is highly localized, with cells near the wound edge experiencing significantly higher normal stresses than shear stresses \cite{bai2023computational}. Furthermore, stress-driven polarization facilitates active coordination between tissue tensions and protrusive and contractile forces \cite{wang2024multiple}.
A critical open question remains the formation of the actin-myosin purse-string via Ca\textsuperscript{2+} signaling. While continuum mechanics models have successfully reproduced some experimental observations of this process, the underlying mechanisms are still not fully understood \cite{roldan2019computational}.}

\subsection*{Hybrid models}
\label{section:hm}

Hybrid models integrate features from multiple approaches to capture the complexity of biological processes across scales \cite{smallwood2009, zheng2005nonlinear, wang2012normal}. In embryonic healing and epithelial gap closure, they combine discrete and continuum approaches, or variations of continuum models, to describe interacting biological processes at multiple levels. This allows for the simulation of cellular and tissue-scale dynamics within a unified framework.

{

Hybrid models typically couple discrete representations of individual cells (e.g., lattice-based, vertex, or particle models) with continuum descriptions of the extracellular matrix (ECM), signaling molecules, or interstitial fluid flow \cite{rejniak2011hybrid, klm2015hybrid}. Examples include cell-based models interacting with a continuous ECM field and discrete cell migration guided by a continuous chemotactic gradient \cite{gonzalez2017hybrid, gonzalez2019agent,hirway2021multicellular, scott2020hybrid}. Hybrid models may also merge stochastic processes (e.g., random cell motility or division, often modelled with Langevin dynamics or Monte Carlo processes) with deterministic continuum equations to capture global dynamics \cite{frederick2001hybrid,bratsun2016chemo,tajvidi2024active}. Here, we consider hybrid models as those that explicitly integrate biochemical kinetics into their mechanical models, rather than relying on coarse-grained or phenomenological couplings. 
}
In hybrid models, biochemical signals are frequently modelled using reaction-diffusion equations, either in continuous space or discretized across cell domains. In a continuum representation, these signals obey partial differential equations of the form:

\begin{equation}
    \frac{\partial c(\vec{\mathbf{x}},t)}{\partial t} = D \vec{\nabla}^2c(\vec{\mathbf{x}},t)+R(c(\vec{\mathbf{x}},t)),
\end{equation}
where $c(\vec{\mathbf{x}},t)$ denotes concentration, $D$ is the diffusion coefficient, and $R(c)$ represents reaction kinetics (like production or decay) \cite{tao2020tuning, bailles2022mechanochemical}. In discrete spatial representation, reaction-diffusion is implemented on a lattice, each node $i$ holds a concentration $c_i(t)$, and the resulting ordinary differential equations follow:

\begin{equation}
    \frac{d c_i}{dt} = \sum_\text{neighbors of i}{D_{ij}(c_j-c_i)}+R(c_i).
\end{equation}
This implementation can also be used at a vertex network scale, where the indexes $i$ correspond to cells, with the neighbors being adjacent cells. $R(c_i)$ in this case represent intracellular reaction kinetics, whilst $D_{ij}$ is a coefficient describing intercellular transport between cells $i$ and $j$. In particular, both implementations can be used simultaneously if both extracellular and intracellular dynamics are being considered \cite{donati2022calcium, yu2024mechanochemical}.

Many hybrid models incorporate chemotaxis, the directed motion of cells in response to chemical signals. In cell-based models, it is usually implemented by biasing individual cell movements according to the local chemical gradient. A minimal implementation, is of the form:
\begin{equation}
    \begin{split}
        \frac{d \vec{\mathbf{r}}_i}{d t} = v_0 \vec{\mathbf{p}}_i + \text{additional terms}\\
        \frac{d \vec{\mathbf{p}}_i}{d t} = -\frac{1}{\tau}\vec{\mathbf{p}}_i + \chi \vec{\mathbf{\nabla}} c(\vec{\mathbf{r}}_i,t),
    \end{split} 
\end{equation}
where $\vec{\mathbf{r}}_i$ is the cell-center position, $\vec{\mathbf{p}}_i$ its polarity, $v_0$ the self-propulsion speed, $\tau$ a persistence time and $\chi$ a chemotactic sensitivy \cite{harris2017simple}. This couples the direction of motion of cells directly to local chemical gradients. In the continuum limit, chemotaxis is usually described using the Keller-Segel system:
\begin{equation}
    \begin{split}
        \frac{\partial n (\vec{\mathbf{r}},t)}{\partial t} &= \vec{\mathbf{\nabla}} \cdot \left(D_n \vec{\mathbf{\nabla}} n(\vec{\mathbf{r}},t) - \chi n(\vec{\mathbf{r}},t) \vec{\mathbf{\nabla}} c(\vec{\mathbf{r}},t)\right)\\
        \frac{\partial c(\vec{\mathbf{r}},t)}{\partial t} &= D_c \vec{\mathbf{\nabla}}^2 c(\vec{\mathbf{r}},t) + R(n(\vec{\mathbf{r}},t), c(\vec{\mathbf{r}},t)),
    \end{split}
\end{equation}
where $n(\vec{\mathbf{r}},t)$ is the cell density, $c(\vec{\mathbf{r}},t)$ is the chemical concentration, $D_n$ and $D_c$ are diffusion coefficients, $\chi$ is the chemotactic sensitivity and $R(n,c)$ represents the reactions kinetics. Cells migrate up the chemical gradient via the flux term $n \vec{\mathbf{\nabla}} c$ \cite{keller1971model, si2012pathway,bubba2020discrete}.

Hybrid models provide a comprehensive framework for studying embryonic healing. These models balance cellular heterogeneity and environmental influences, effectively bridging discrete and continuous processes. While the underlying equations, such as elasticity, chemotaxis, or agent-based motility, are often drawn from established models, hybrid approaches are distinguished by how they couple these components across scales. This typically comes at the cost of increased computational burden and parameter complexity, especially in defining cross-scale couplings. These challenges, which may limit model interpretability and scalability, are generally acknowledged in the field. 

{In recent applications of hybrid models, mechanical feedback and Ca\textsuperscript{2+} signaling emerge as recurring mechanisms coordinating collective cell behavior. Several studies have explored Ca\textsuperscript{2+} dynamics, illustrating how mechanical properties and cellular anisotropy alter the propagation of Ca\textsuperscript{2+} waves, having a directional preference \cite{narciso2015}, and how wound-induced microtears and cell loss trigger distinct Ca\textsuperscript{2+} entry mechanisms \cite{shannon2017}. Other studies link extracellular ATP release to enhanced collective migration through Ca\textsuperscript{2+} influx \cite{odagiri2022} and highlight intercellular IP3 transfer as essential for coordinating Ca\textsuperscript{2+} signals across distal cells \cite{stevens2023mathematical}. Other hybrid approaches have examined migration and mechanical feedback. Contact inhibition and substrate interactions were shown to drive local relaxation and collective reorganization after injury \cite{coburn2018role}, while mechanochemical feedback through ERK signaling facilitated long-range guidance cues during cell migration to close the wound \cite{hino2020erk}. Voronoi-based and finite-element models successfully integrated cell- and tissue-scale mechanics to predict durotaxis and gap closure \cite{gonzalezvalverde2018}. Cell motility waves were found to depend on contractility and cortical tension, influencing edge stability \cite{khataee2020multiscale}. Hybrid models have also enhanced our understanding of epithelial-mesenchymal transition (EMT) in embryonic healing. A coupled CPM-finite element framework predicted EMT initiation at tissue boundaries due to mechanochemical interactions \cite{hirway2021multicellular}, while a multiscale model incorporating YAP signaling revealed conditions for fingering instabilities \cite{mukhtar2022multiscale}. Computational models demonstrated that ECM composition modulates EMT-driven closure, independent of TGF-$\beta$1 signaling \cite{o2021proteolytic}. Together, these studies demonstrate the power of hybrid modeling in capturing the interplay between biochemical and mechanical processes during wound repair.}

\subsection*{Data-driven Models}
\label{section:data}

Data-driven models leverage experimental and imaging data to construct predictive frameworks for embryonic healing dynamics. Unlike traditional physics-based models that rely on predefined laws, data-driven methods identify patterns, correlations, and features directly from empirical observation. Put in more concrete terms, while the previously discussed models would have given sets of predefined parameters to predict the observed healing behavior, data-driven models would use experimentally observed healing processes to estimate the relevant parameters. These models often employ statistical methods, machine learning, or hybrid techniques that integrate empirical observations with physical modeling \cite{stillman2023generative, battula2020medical}.

Recent advances in experimental imaging and computing power have enabled the acquisition of high-resolution spatiotemporal data on cell and tissue dynamics. As a result, data-driven models have gained traction for capturing complex behaviors that may be difficult to express with conventional physical models \cite{schwayer2023connecting, bruckner2024learning}.

Data-driven methods in wound modelling split broadly into several categories, each ground in distinct principles. One common approach envolves the use of linear or nonlinear regression techniques to identify trends in the experimental data. These methods are particularly useful when the goal is to characterize the healing using minimal assumptions \cite{watson2024}.  A typical linear regression model has a dependent variable,  an $M$-dimensional vector $\hat{\mathbf{y}}$, where $M$ is the number of samples and a collection of $N$ independent variables $\mathbf{X} = \left[\hat{\mathbf{1}} ,\hat{\mathbf{x}}_1,...,\hat{\mathbf{x}}_N\right]$, where each $\hat{\mathbf{x}}_i$ is a $M$-dimensional vector and $\hat{\mathbf{1}}$ is a vector in which all coefficients are equal to 1. The predicted value is given by:

\begin{equation}
    \hat{\mathbf{y}} =  \mathbf{X} \cdot \beta+\hat{\mathbf{\epsilon}},
\end{equation}where $\beta$ is an $N+1$ dimensional parameter vector and $\hat{\epsilon}$ is the $M$-dimensional prediction error. The linear fit requires changing the values of $\beta$ such that $\hat{\epsilon}$ is minimized. This is usually done by the least-squares method, where the loss function, defined as $\mathcal{L}= \sum_{i= 1}^{M}{\epsilon}_i^2 = \sum_{i= 1}^{M}(\hat{\mathbf{y}}-\mathbf{X} \beta)_i^2$, is minimized:
\begin{equation}
    \frac{\partial \mathcal{L}}{\partial \beta_i} = 0,
\end{equation} where the loss function is minimized for all components $\beta_i$. This minimization process is functionally how most statistical methods operated, where what is being changed are not the variables but the parameters for the fixed observed variables \cite{su2012linear}. 

A more sophisticated class of models employ Bayesian inference to incorporate prior biological knowledge, to rigorously quantify uncertainty in the predictions the model makes. These methods are valuable in contexts where embryonic healing parameters may vary significantly across systems \cite{kachouie2006probabilistic, metzner2015superstatistical,zanca2022}.  An aspect of Bayesian inference is the use of Bayes theorem:

\begin{equation}
    p(\theta|\mathbf{X}) = \frac{p(\mathbf{X}|\theta) p(\theta)}{p(\mathbf{X})},
\end{equation} where $\theta$ are the model parameters, $\mathbf{X}$ is the observed data, $p(\theta)$ is the prior probability distribution, which is our initial assumption about the distribution of parameters $\theta$ without considering the observed data, $p(\mathbf{X}|\theta)$ is the probability of observing the data $\mathbf{X}$ given the parameters $\theta$, or the likelihood function, $p(\mathbf{X})$ is the data distribution, and $p(\theta|\mathbf{X})$ is what is called the posterior distribution, the distribution we are trying to estimate \cite{turkman2019, vandeschoot2021}.   Hybrid data-driven models integrate empirical data directly into traditional physical models to refine parameter estimates and improve predictions \cite{rodrigues2021deeper, brunton2019methods, martinaperez2022}. 

In parallel, machine learning has emerged as a versatile tool in analyzing high-dimensional embryonic healing data. These techniques ranging from supervised, unsupervised, and deep learning, have been increasingly applied to classify wound stages, extract features, and predict embryonic healing outcomes from imaging data \cite{cichos2020machine, le2023unveiling}. For instance, classification models might output wound closure rates, wound edge velocity fields, or epithelial sheet strain maps.

{These approaches have been applied in a variety of ways. For example, early studies used confocal microscopy image analysis and automated image segmentation on time-lapse sequences to predict the role of specific chemicals in promoting wound closure \cite{zulueta2014automated}. Other work has combined information on cell density and heterogeneity to show that wounds close more slowly when there is low heterogeneity in cell migration modes \cite{vishwakarma2020dynamic}. Methods such as proper orthogonal decomposition have been used to determine the direction of collective cell motion during migration \cite{han2022proper}. Furthermore, machine learning has been applied using dynamical systems-inspired approaches to extract pathways leading to epithelial-to-mesenchymal transition \cite{hu2023dynamical}, as well as variational autoencoders to capture hierarchies of cell behaviors during wound closure, quantified through area reduction and distance metrics \cite{backova2023modeling, turley2024deep}. Bayesian inference methods have also proven useful for parameter estimation in complex embryonic healing models \cite{ariza2024bayesian}.}

Despite their growing popularity, these models present several challenges. Due to their reliance on experimental data, these models may capture behaviors that fail to generalize across varying conditions or datasets \cite{recht2019imagenet, shankar2021image}. Machine learning methods, while increasingly prevalent, often function as black boxes, limiting interpretability and making it difficult to infer the underlying mechanics of wound closure \cite{kadambi2023incorporating}. 

\section*{Discussion}
\label{section:discussion}

{Throughout this review, we have examined a range of physical models that describe embryonic healing, spanning cell-based, continuum and hybrid formulations, each at distinct spatial and temporal scales. Many of these studies have successfully reproduced embryonic wound closure assays \textit{in silico} without setting the outcomes in advance \cite{wortel2021local, ton2024mechanical, najem2016phase, arciero2013using, ravasio2015regulation}.  In this section, we discuss to what extent have current models addressed the core challenges identified in earlier embryonic healing reviews and where can the existing models be used more effectively, or strategically extended.

{ The models considered in this review can be classified by how tissue behavior emerges. Cell-based models are rule-based, with defined behaviors for the individual cells that lead to emergent properties at the tissue level. Continuum models can be phenomenological to varying degrees, depending on the framework. Viscoelastic models and the standard phase field models are phenomenological, as the properties expected are directly incorporated into the energy functional. In contrast, hydrodynamic models are derived from physical principles, and it is expected that tissues match their behavior in some conditions.}


Embryonic healing involves nonlinear, time-dependent tissue mechanics and spatial anisotropy that are tightly regulated by chemical signals and active cellular behavior. Each model captures only a subset of these coupled dynamics. Although cell-based and continuum models each address complementary aspects of tissue mechanics,  this fragmentation has made it difficult to form a coherent, multiscale theory of embryonic wound closure \cite{jorgensen2016mathematical, weihs2016review}. 

Zulueta-Coarasa \textit{et al}  identified five core mechanical features driving embryonic wound repair: collective cell motion, coordination by junctional and cytoskeletal rearrangements, the purse-string contraction modulated by Ca2+ and protrusive activity \cite{zulueta2017tension}. These features form the backbone of any adequate model. Building on this, the wound geometry, polarization, ECM interactions, and general signaling pathways must also be included to capture the broader properties of the model. Table \ref{tab:table2} summarizes how each model has addressed these key features, offering a comparative overview of the state-of-the art, as of 2024.

\begin{table}[h!]
    \footnotesize
    \centering
    \hspace{-2in}
    \begin{tabular}{|m{8em}|m{1.2cm}|m{1.2cm}|m{1.2cm}|m{1.5cm}|m{1.3cm}|m{1.2cm}|m{1.0cm}|}\hline
 & \multicolumn{2}{|c|}{\textbf{Cell-based (CB)}}& \multicolumn{3}{|c|}{\textbf{Continuum (CM)}}& \multicolumn{1}{|c|}{\textbf{Hybrid (HM)}}&\multicolumn{1}{|c|}{\textbf{Data-driven (DD)}}\\\hline
         \textbf{Feature}&  \textit{Lattice}&  \textit{Network}&  \textit{Phase-Field}&  \textit{Hydro-dynamic}&  \textit{Visco-elastic} &  & \\\hline

\multicolumn{8}{|c|}{\textbf{Explicit wound-boundary modelling}} \\ \hline
         Purse-string contractility and protrusive activity&  \cite{noppe2015}; *\cite{khataee2020multiscale}&  \cite{staddon2018,zulueta2018, ajeti2019, babu2024interplay}&  \cite{feng2023physical, pozzi2024geometric}&  \cite{ arciero2013using, zhao2023analytical, chen2019}&  \cite{wyczalkowski2013computational, brugues2014, lee2011crawling,ravasio2015, he2019curvature}&  \cite{khataee2020multiscale}& \\\hline
 Wound geometry& & \cite{staddon2018, babu2024interplay}& \cite{feng2023physical, pozzi2024geometric}& \cite{arciero2013using, zhao2023analytical}& \cite{vedula2015, ravasio2015, he2019curvature}& &\\\hline\hline

\multicolumn{8}{|c|}{\textbf{General epithelial mechanics and signaling}} \\ \hline
  Cell mechanical behavior& & \cite{tetley2019, mosaffa2020, ton2024mechanical}; *\cite{gonzalezvalverde2018}& & \cite{berlyand2024bifurcation}& \cite{cochet2014,carvalho2021, bai2023computational}; *\cite{gonzalezvalverde2018}& \cite{gonzalezvalverde2018}&\\\hline
         Cytoskeletal and junctional remodelling&  &  \cite{tetley2019, mosaffa2020}&  &  \cite{chen2019}&  \cite{vedula2015, wei2020actin}&  & \\\hline
 Cellular migration and EMT& \cite{scianna2012cellular, scianna2015, wortel2021local, roy2021intermediate}; *\cite{coburn2018role, hino2020erk, hirway2021multicellular}& \cite{zanca2022push}& \cite{palmieri2015multiple, najem2016phase}& \cite{arciero2011, berlyand2024bifurcation}& *\cite{hirway2021multicellular}& \cite{coburn2018role, hino2020erk, hirway2021multicellular,mukhtar2022multiscale}&\cite{vishwakarma2020dynamic, han2022proper, hu2023dynamical}\\\hline
         ECM and its mechanical properties&  \cite{scianna2012cellular, scianna2015, roy2021intermediate}; $*$\cite{coburn2018role}&  \cite{staddon2018, ajeti2019}&  &  &  &  \cite{coburn2018role, o2021proteolytic}& \\\hline
         Cell polarity&  \cite{wortel2021local}&  &  &  \cite{he2020theoretical}&  \cite{wang2024multiple}&  & \\\hline
         Chemical signaling&  *\cite{odagiri2022}&  *\cite{stevens2023mathematical}&  &  \cite{asakura2021hierarchical}&  &  \cite{odagiri2022, stevens2023mathematical}& \cite{zulueta2014automated}\\\hline
         Mechano-chemical feedback&  *\cite{hino2020erk, hirway2021multicellular, mukhtar2022multiscale}&  &  &  \cite{chen2024piezo1}&  &  \cite{narciso2015, shannon2017, hino2020erk, hirway2021multicellular, mukhtar2022multiscale, o2021proteolytic}& \\\hline
         Parameter inference&  &  &  &  &  &  & \cite{ariza2024bayesian}\\ \hline
    \end{tabular}
    \caption{Several models and the main references where core features of embryonic healing are addressed. Asterisks (*) denote hybrid models and indicate when a model serves as a basis for the hybridization. The table is split for interpretative clarity: the top block highlights features that explicitly emphasize wound-boundary dynamics, while the bottom block covers features from general epithelial mechanics and signaling relevant to wound healing.}
    \label{tab:table2}
\end{table}


Evidence across both cell-based and continuum models confirms that efficient embryonic wound closure requires the dynamic coordination between actomyosin purse-string contraction and lamellipodial crawling. Vertex and lattice models show that actomyosin cables dominates in small wounds, while protrusive activity becomes dominant as gaps widen, a switch controlled by adhesiveness,  contractility and local curvature \cite{noppe2015, ravasio2015, staddon2018, ajeti2019, tetley2019,zhao2023analytical}.  Network models further capture spatial heterogeneity and anisotropic force distributions around irregular wounds, to promote more efficient closure \cite{zulueta2018}, a feature harder to realize in purely continuum models.  Notably, several models reproduce the experimentally observed curvature-dependent mode-switch  \cite{lee2011crawling, ravasio2015, staddon2018, he2019curvature, chen2019}.  Lattice models, however struggle to resolve fine-grained geometric features beyond areas and perimeters, limiting their ability to predict curvature-driven feedback on actomyosin dynamics.


Models across scales consistently highlight the central role of collective cell migration in embryonic healing, with both the extracellular matrix and cell polarity emerging as key controllers of migratory behavior. Experimental studies have showed that cells shift from apical–basal to front–rear polarity to effectively migrate and close wounds \cite{chapnick2014, poujade2007, Begnaud2016}. Discrete models that combine discrete cell behaviors and tissue-level mechanics consistently show that collective cell migration is regulated by changes in cell polarity, often coupled to epithelial-to-mesenchymal transition (EMT). Hybrid models further suggest that EMT, initiated at tissue boundaries via mechanochemical feedback, not only enhances closure, but can also produce fingering instabilities \cite{hirway2021multicellular, mukhtar2022multiscale, o2021proteolytic, zhao2023analytical}.

Simultaneously, lattice and hybrid models demonstrate that the physical properties of the ECM, including fiber density, adhesiveness, and orientation, strongly influence both speed and direction of migrating sheets \cite{scianna2012cellular, scianna2015}. ECM remodeling, mediated by integrins and matrix proteins such as fibronectin and collagen, further modulates durotaxis and the tissue's viscoelastic response \cite{gonzalezvalverde2018, vedula2014, ravasio2015regulation, ajeti2019}. 

Given that polarity and ECM interactions fundamentally govern migration modes, it is critical to identify which models incorporate these features, and which do not. In the works we considered, continuum models typically neglected explicit ECM structure, whereas lattice and network schemes incorporate them. Polarity mechanisms are absent from basic network models, but appear naturally in active polar or nematic models. Recent Voronoi and active-vertex models naturally include polarity and alignment rules \cite{barton2017, staddon2022interplay}. Although phase-field models here lack both ECM and polarity mechanisms, there are recent examples of both being  successfully integrated in their models \cite{zhang2020active, chiang2024multiphase}. Likewise, continuum approaches can and do include elastic or viscoelastic substrates in more advanced formulations \cite{garciagonzalez2020computational,plan2020active, adar2022activegel}.  In particular, an avenue for further exploration is to model the feedback loop that can existing between tissue deformation and ECM deformation during closure, and how this may impact its efficiency.


The remodelling of the actin-myosin network within the cells, and the dynamics of adherens junctions are also prevalent across most models, underscoring their role in transmitting forces during collective epithelial migration. Continuum models often introduce junctional tension through phenomenological stress terms, whereas vertex and hybrid formulations can natively encode these interactions at each cell-cell interface \cite{hunter2017, carvalho2018, rothenberg2019}.  Lattice and phase-field approaches, by contrast, rarely feature junctional mechanics, but can incorporate adhesive interactions.

Beyond acto-myosin contractility, other { cytoskeletal elements, particularly microtubules and intermediate filaments}, remain largely underrepresented in current models, despite their known roles in cell polarity and intracellular transport \cite{hunter2017, rothenberg2019}. Additionally, the influence of the nucleus must be considered as it is not often considered in wound healing models. Incorporating these cytoskeletal components is not just an additional factor, but it is essential for building a comprehensive mechanochemical model of embryonic wound repair. 


Chemical signaling and mechanochemical feedback have proven crucial for capturing the full complexity of embryonic healing. Hybrid models show that Ca\textsuperscript{2+} waves, whose propagation depends on mechanical tension and cell geometry, drive localized actin remodeling \cite{narciso2015, shannon2017}. These signals, enhanced by ATP and IP\textsubscript{3}, synchronize collective movements across the epithelial sheet \cite{odagiri2022, stevens2023mathematical}. Beyond hybrid models, active polar and nematic models show that mechanical stress fields coupled with these chemical signals can bias cell polarity and even trigger instabilities at the wound front \cite{he2020theoretical, zhao2023analytical, berlyand2024bifurcation}. Continuum models more readily incorporate reaction-diffusion equations to capture the chemical signals, though lattice or network models increasingly incorporate chemical signaling through hybrid implementations.

Menon \textit{et al} highlighted a persistent gap in wound-healing modeling which was the near-complete lack of statistical approaches applied for parameter estimation, which hinders the ability to interpret theoretical and computational results with experimental observations\cite{menon2021mathematical}. In parallel, recent advances in computer vision and deep learning now appear as promising tools for extracting high-dimensional features, like cell trajectories or fluorescent chemical signals, directly from experimental images \cite{turley2022good, lim2024forced}. Current applications have focused on characterizing collective migration patterns and extracting complex chemical signals, to determine how specific chemical factors drive closure dynamics \cite{zulueta2014automated, han2022proper, hu2023dynamical}. This ability to analyze large, unstructured datasets for hidden correlations represents an opportunity for data-driven approaches within embryonic wound repair.

Despite recent progress, inferring mechanochemical coupling constants, such as those linking strain to signaling activation, remains a central challenge. The combination of model nonlinearity and noisy experimental data has so far made it difficult to obtain estimates that are both reliable and generalizable \cite{menon2021mathematical}. Yet, advances in deep learning and Bayesian inference methods suggest that data-driven estimation of these parameters is within reach, bringing us closer to genuinely predictive, physics-informed models \cite{ariza2024bayesian}.  


There is no shortage of modeling efforts addressing key features of embryonic wound repair, yet important gaps persist within certain frameworks. Lattice models, for example, offer fine‑grained control of subcellular mechanics but struggle to represent evolving wound geometry and lack true dynamics, whereas phase‑field schemes capture smooth cell-boundary deformations yet omit explicit junctional and cytoskeletal dynamics. Hydrodynamic approaches readily incorporate polarity fields, chemical gradients, and large‑scale flow, making them ideal for wounds much larger than a single cell, but their continuum assumptions wash out the heterogeneity and anisotropy intrinsic to real tissues. Recognizing these trade‑offs reveals opportunities for extending each framework: lattice models could adopt curvature‑sensitive rules to better handle shape, phase fields might integrate discrete junctional forces, and hydrodynamic theories can borrow heterogeneity from network approaches.

{ The choice of the model also depends on the intended purpose. Simpler models with fewer equations or parameters allow general principles to be inferred and are more interpretable, but may be less quantitatively accurate. More detailed models can reproduce observed tissue behavior more faithfully, though at the cost of interpretability. The appropriate framework depends on whether the goal is to understand underlying mechanisms or to closely match experimental observations.

Depending on the biological scale and processes of interest, different models are appropriate. Where individual cell morphologies are relevant, lattice or multiphase field models are adequate. Vertex models are well suited for junctional mechanics, while hydrodynamics or viscoelastic models capture tissue-scale collective behavior. Wound geometry also affects which mechanisms dominate: purse-string forces are negligible in scratch assays, whereas crawling is less prominent in non-adherent substrates. This can guide model selection for experimentalists studying epithelial wound healing.}

Among the modeling frameworks reviewed, the vertex model stands out in its ability to capture a broad spectrum of behaviors relevant to embryonic healing. By representing each cell as a deformable polygon with explicit edges and vertices, it naturally encodes heterogeneity, anisotropic tension, junctional remodeling, and polarity alignment, all within a single, easily implemented scheme. Vertex models excel at capturing the balance between purse‑string and protrusive forces across a wide range of wound sizes. While they are less effective at modelling mesenchymal cell behavior, and miss sub-pixel boundary smoothness at the smallest scales, they nonetheless cover more of the essential ingredients of embryonic epithelial repair than any other single paradigm. Researchers should select a model based on the scale and mechanisms most relevant to their system. However, when those are uncertain, the vertex model offers a flexible and comprehensive default starting point.

{A broader limitation of the current modeling landscape is the limited cross-validation across models. Although several mechanisms, including the cooperation between purse-string contraction and protrusive crawling, have been reproduced qualitatively by different approaches, systematic quantitative comparisons remain rare. Models that are designed to represent similar epithelial sheets often rely on distinct assumptions, coarse-graining strategies and parametrizations, leading to quantitatively different predictions even when they capture the same phenomenology. As a result, the predictive power and generalizability of any single model remain difficult to assess in the absence of common benchmarks or formal parameter mappings.}

{Other forms of tissue repair, like single-cell extrusions, which occur in three dimensions, or transient junctional tears, constitute the most common wounds in embryonic epithelia. These wounds are treated as topological defects rather than free boundaries in a plane, as in the models considered here. While biologically relevant, these processes fall outside the scope of this review, which focuses on wounds dominated by collective, multicellular dynamics. There are, however, some recent examples of physical models applied to study cell-extrusions and junctional tears \cite{barton2017, saw2017topological, okuda2020a, sonam2023mechanical}. Although these processes are physically distinct from multicellular free-boundary wounds, they share underlying mechanisms, which shows the broad range of epithelial repair mechanisms.

While this review focuses on embryonic epithelial healing, many of the physical results obtained from these models are relevant to adult wound repair, provided the underlying assumptions are reconsidered carefully. Model implementations related to collective cell migration, proliferation and mechanochemical feedback are largely transferable, although their results must be re-evaluated in three dimensions. In particular, reaction-diffusion models for signaling, or active matter models to describe polarity can be extended to adult healing with some formal changes. 

The major differences are in tissue architecture and mechanical environment. While embryonic healing is three dimensional, it can often be well described by two dimensional models as we have seen across this review. Adult wound healing however occurs in a three dimensional extracellular matrix populated by fibroblasts and being actively remodelled. In contrast, in the embryonic models reviewed here, ECM is a quasi-two-dimensional substrate supporting a confluent epithelial sheet. ECM is a material that experiences plastic deformations and fiber realignment, and plays a far more active role in adult wound healing. Thus the models which include ECM as an elastic or passive surface in embryonic healing contexts are not directly applicable. 

Additional processes central in adult healing, such as inflammation and angiogenesis, introduce new active components and length scales. While these processes lie outside the scope of the present review, it is notable that several of the models discussed here, have already been adapted to describe angiogenic growth and vascular remodelling, where three dimensional geometry is essential and cannot be abstracted away \cite{menon2021mathematical, flegg2020current}.

}

Finally, drawing parallels from morphogenesis models may help resolve the remaining challenges in polarity, cellular organization, and mechanochemical feedback \cite{recho2015mechanics, tlili2018collective, bogdan2018fingering, perez2019active, staddon2019mechanosensitive}. The diversity of models reviewed here, each built on distinct physical assumptions, nonetheless converge on similar biological observations. Highlighting that hybridization between models remain the most promising approach. 


Looking ahead, we expect hybrid models to become increasingly central, particularly in integrating ECM-tissue mechanics, incorporating underexplored cytosketelal elements and chemical signaling pathways. 
As hybrid models become more complex, the need for data-driven parameter inference will become more acute. Their greatest limitation, the lack of robust coupling parameters, could be overcome through tighter integration with experimental data and machine learning pipelines. { To contextualize these modeling directions, Table~\ref{tab:table3} summarizes the experimental systems, species, wounding assays, and open-source implementations associated with the models reviewed. The overview is not exhaustive; the absence of specific assays or biological systems reflects the current literature and not intrinsic limitation of the models. It represents an opportunity for future exploration.}

\begin{table}[h!]
    \centering
    \footnotesize
    \hspace{-1in}
    \begin{tabular}{|l|>{\raggedright\arraybackslash}p{0.25\linewidth}|>{\raggedright\arraybackslash}p{0.2\linewidth}|>{\raggedright\arraybackslash}p{0.2\linewidth}|>{\raggedright\arraybackslash}p{0.2\linewidth}|}\hline
         \textbf{Model} & \textbf{Biological System} & \textbf{Species} & \textbf{Wounding Assay} & \textbf{Open-source implementations} \\ \hline \hline
         
         \textit{Lattice} & Epithelial monolayer (in vitro) & Human (OVCAR-3, MCF-7); Canine (MDCK) & Scratch assay; Laser ablation; Barrier removal assay & CompuCell3D; Morpheus \\ \hline
         
         \textit{Network} & Embryonic epithelial monolayer (in vivo); embryonic-like monolayer (in vitro) & Canine (MDCK); Drosophila embryo & Laser ablation; Barrier removal assay & SurfaceEvolver; SAMoS; cellGPU \\ \hline
         
         \textit{Phase-field} & Embryonic-like epithelial monolayer (in vitro) & Canine (MDCK) & Scratch assay; Barrier removal assay & Finite element solvers (FEniCS, FreeFEM) \\ \hline
         
         \textit{Hydrodynamic} & Epithelial monolayer (in vitro) & Canine (MDCK); Human keratinocytes & Scratch assay; Barrier removal assay & Finite element solvers (FEniCS, FreeFEM) \\ \hline
         
         \textit{Viscoelastic} & Embryonic epithelial tissue (in vivo); epithelial monolayer (in vitro) & Canine (MDCK); Human keratinocytes; Chick embryo & Incision; Barrier removal assay & Finite element solvers (FEniCS, FreeFEM) \\ \hline
         
    \end{tabular}
    \caption{Summary of the biological systems, species, wounding assays, and open-source implementations for the main physical models of epithelial wound healing reviewed. The table distinguishes between in vivo embryonic tissue and in vitro monolayer systems, highlighting where the reviewed models have been applied. Examples listed are not exhaustive. Hybrid and Data-driven systems are not included, as their application is not restricted to specific systems.}
    \label{tab:table3}
\end{table}

While individual models offer focused results on specific aspects, the convergence of findings across different methods underscores a broader principle: mechanical forces and chemical signaling are intrinsically linked in embryonic healing. { At present, however, this convergence is primarily assessed qualitatively, based on whether models reproduce similar experimentally observed behaviors.} Models that successfully integrate both, particularly through hybrid or data-driven approaches, are best positioned to advance the field.  A more systematic comparison would require deriving effective continuum parameters from cell-based models, enabling quantitative links between microscopic rules and macroscopic material properties.
{Such a formal theoretical translation between models is largely absent at present and constitutes a central open challenge for the modeling of embryonic wound healing.}}
\newpage
\begin{mdframed}[linewidth=1pt, linecolor=black, roundcorner=5pt, innerleftmargin=10pt, innerrightmargin=10pt]
{\bf Key Summary Points:}
\begin{itemize}[leftmargin=*]
    \item Multiple physical models have been applied to embryonic wound healing, with distinct physical assumptions;
    \item Different models specialize in different tissue features, with discrete models capturing cell-level mechanics and continuum models capturing tissue scale flows and chemical fields;
    \item Appropriate model choice should match the biological and experimental question of interest;
    \item Quantitative validation is limited, highlighting challenges in parameter inference and the need for data-driven integration across models.
\end{itemize}

{\bf Key Papers:}
\begin{itemize}[leftmargin=*]
    \item Alert R, Trepat X. Physical models of collective cell migration. Annu Rev Condens Matter Phys. 2020;11:77--101.;
    \item Begnaud S, Chen T, Delacour D, Mège R, Ladoux B. Mechanics of epithelial tissues during gap closures. Curr Opin Cell Biol. 2016;42:52--62;
    \item Rothenberg KE, Fernandez-Gonzalez R. Forceful closure: cytoskeletal networks in embryonic wound repair. Mol Biol Cell. 2019;30(12):1353--1358.;
    \item Zulueta-Coarasa T, Fernandez-Gonzalez R. Tension (re) builds: biophysical mechanisms of embryonic wound repair.  Mech Dev. 2017;144:43--52.;
    \item Fletcher AG, Osborne JM. Seven challenges in the multiscale modeling of multicellular tissues. Wires Mech Dis. 2022;14(1):e1527.
\end{itemize}
\end{mdframed}



\begin{thebibliography}{100}
\bibitem{haensel2018}
Haensel D, Dai X.
\newblock Epithelial-to-mesenchymal transition in cutaneous wound healing: where we are and where we are heading.
\newblock Dev Dyn. 2018;247(3):473--480.
\end{thebibliography}
\end{document}